\documentclass{article}

\usepackage[preprint]{neurips_2025}

\usepackage[utf8]{inputenc} 
\usepackage[T1]{fontenc}    
\usepackage{hyperref}       
\usepackage{url}            
\usepackage{booktabs}       
\usepackage{amsfonts}       
\usepackage{nicefrac}       
\usepackage{microtype}      
\usepackage[table]{xcolor}  
\usepackage{graphicx}
\usepackage{multirow}
\usepackage{caption}
\usepackage{enumitem}
\usepackage{xspace}
\usepackage{wrapfig}
\usepackage{supertabular}
\usepackage{tcolorbox}

\newcommand{\benchmark}{SWE-bench-Live\xspace}
\newcommand{\method}{\textsc{RepoLaunch}\xspace}

\newcommand{\logo}{\includegraphics[height=0.8em]{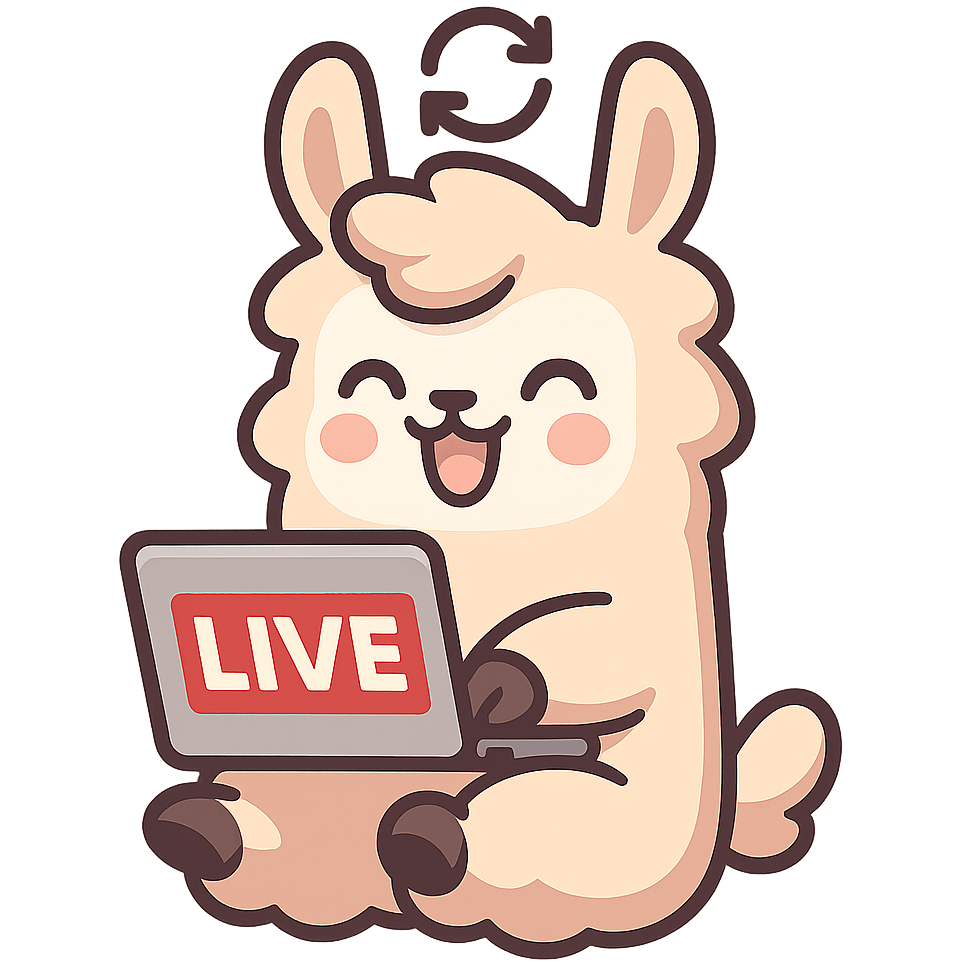}}

\title{SWE-bench Goes Live!}

\author{%
Linghao Zhang \quad Shilin He \quad Chaoyun Zhang \quad Yu Kang \quad Bowen Li \quad \\
\textbf{Chengxing Xie} \quad \textbf{Junhao Wang} \quad \textbf{Maoquan Wang} \quad \textbf{Yufan Huang} \quad  \textbf{Shengyu Fu} \\ 
\textbf{Elsie Nallipogu} \quad \textbf{Qingwei Lin} \quad \textbf{Yingnong Dang} \quad
\textbf{Saravan Rajmohan} \quad \textbf{Dongmei Zhang} \\
Microsoft
}

\begin{document}

\maketitle

\vspace{-2em}
\begin{center}
    \href{https://swe-bench-live.github.io/}{\logo{} \textcolor{blue}{\texttt{Leaderboard}}} \quad
    \href{https://github.com/microsoft/SWE-bench-Live}{\includegraphics[height=0.8em]{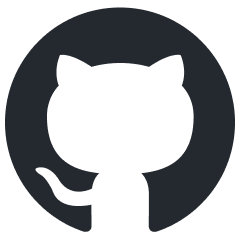} \textcolor{blue}{\texttt{GitHub}}} \quad
    \href{https://huggingface.co/SWE-bench-Live}{\includegraphics[height=0.8em]{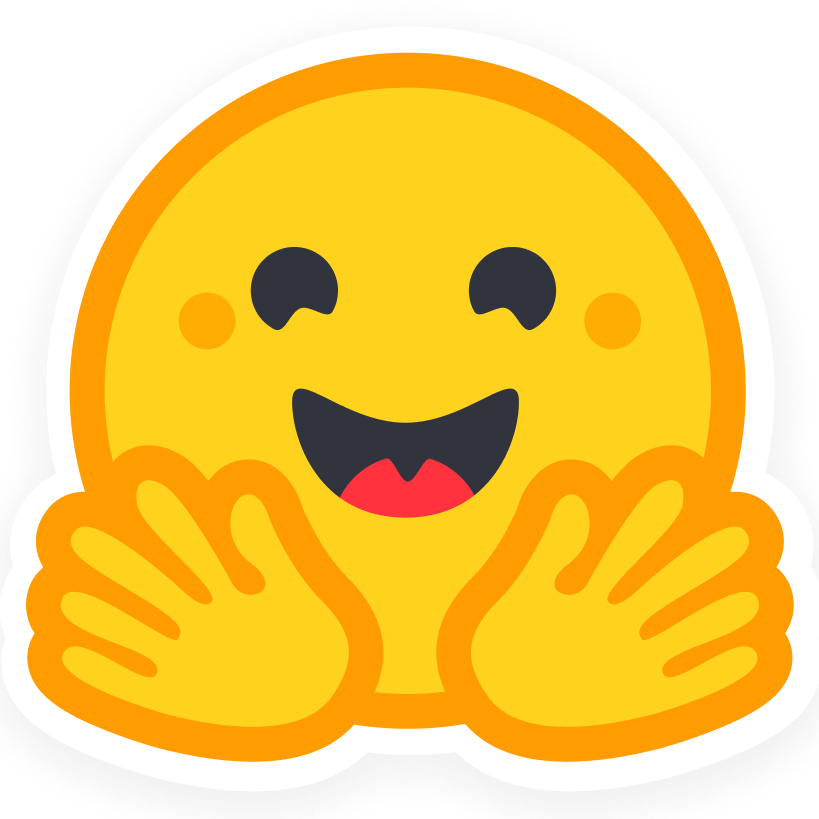} \textcolor{blue}{\texttt{HuggingFace}}} \quad
\end{center}

\begin{abstract}
    The issue-resolving task, where a model generates patches to fix real-world bugs, has emerged as a key benchmark for evaluating the capabilities of large language models (LLMs). While SWE-bench has become the dominant benchmark in this domain, it suffers from several limitations: it has not been updated since its release, is restricted to only 12 repositories, and relies heavily on manual effort for constructing test instances and setting up executable environments, significantly limiting its scalability. We present \textbf{SWE-bench-Live}\footnote{Homepage of SWE-bench-Live: \url{https://swe-bench-live.github.io/}, Code at \url{https://github.com/SWE-bench-Live}, and Dataset at \url{https://huggingface.co/SWE-bench-Live}}, a \textit{live-updatable} benchmark designed to address these limitations. Our initial release includes 1,319 tasks derived from real GitHub issues created since 2024, spanning 93 repositories. Each task is accompanied by a dedicated Docker image to ensure reproducible execution. Additionally, we introduce an automated curation pipeline that streamlines the entire process from instance creation to environment setup, removing manual bottlenecks and enabling scalability and continuous updates.
    We evaluate a range of state-of-the-art models and agent frameworks on SWE-bench-Live, offering detailed empirical insights into their real-world bug-fixing capabilities. By providing a fresh, diverse, and executable benchmark grounded in live repository activity, SWE-bench-Live supports reliable, large-scale assessment of code LLMs and code agents in realistic development settings.
\end{abstract}

\section{Introduction}
Large language models (LLMs) have fundamentally reshaped the landscape of software engineering \cite{fan2023large}, powering tools such as Cursor \cite{cursor2025} and GitHub Copilot \cite{dakhel2023github} that are now integral to modern development workflows. 
These models have transformed key stages of the software development lifecycle—automated code generation, bug detection, and issue resolution—leading to substantial gains in developer productivity. 
To systematically assess LLM capabilities across these tasks, a variety of curated benchmarks have been developed, including HumanEval \cite{chen2021evaluating}, MBPP \cite{austin2021program}, SWE-bench \cite{swebench}, DI-Bench \cite{zhang2025di}, and OpenRCA \cite{xuopenrca}. 
These benchmarks are instrumental in identifying both the strengths and limitations of LLMs in diverse programming and maintenance settings.

Among them, SWE-bench~\cite{swebench} and its variants, such as Multimodal SWE-bench~\cite{yang2024swebenchmultimodalaisystems} and Multi-SWE-bench~\cite{zan2025multi}, have become standard for evaluating LLMs on the issue resolution task, where models are required to comprehend complex codebases, interact with execution environments, and generate patches that fix real-world issues.
However, as LLMs evolve rapidly, existing benchmarks exhibit several critical limitations that undermine their continued utility:

\begin{figure}[!t]
    \centering
    \includegraphics[width=\linewidth]{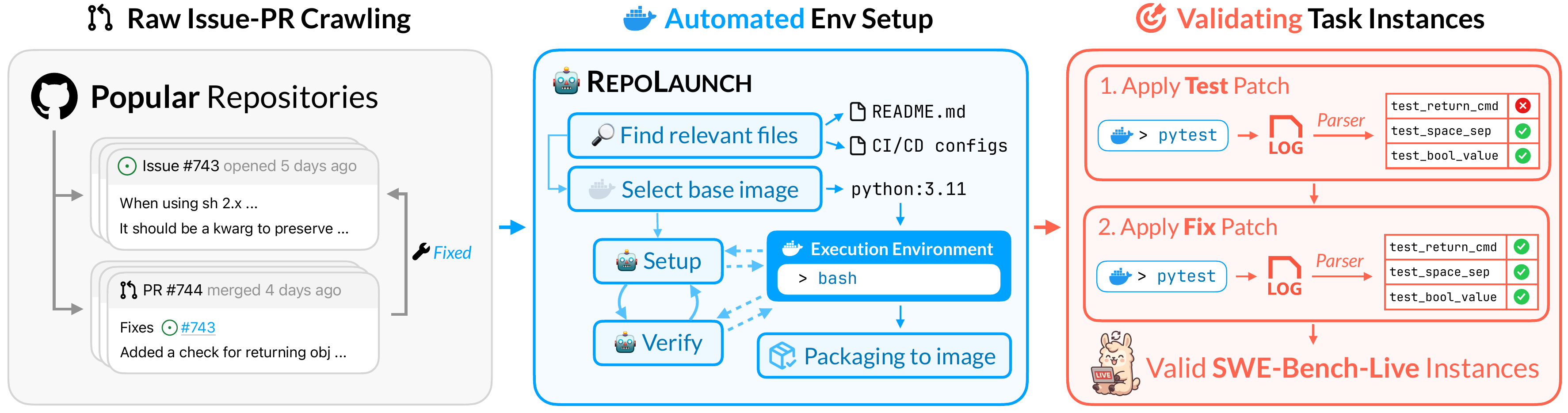}
    \caption{The automatic construction pipeline of \benchmark.}
    \label{fig:overview}
\end{figure}

\begin{table}[ht]
    \centering
     \vspace{-1em}
    \caption{Comparison with existing issue resolving benchmarks.}
    \resizebox{\textwidth}{!}{
    \begin{tabular}{c|ccccc}
    \toprule
      \textbf{Dataset} & \textbf{Date}  & \textbf{\#Instances} & \textbf{\#Repository} & \textbf{Real/Synthetic} & \textbf{Curation}\\
      \midrule
      SWE-bench~\cite{swebench} & Oct, 2023 & 2,294 & 12 & Real & Manual \\
      SWE-bench-Verified~\cite{swebench-verified} & Aug, 2024 &  500 & 12 & Real &  Manual \\
      SWE-Gym~\cite{swegym} & Dec, 2024 & 2,438 & 11 & Real & Manual \\
      Multi-SWE-bench~\cite{zan2025multi} & Apr, 2025 & 1,632 & 39 & Real & Manual \\
      SWE-smith~\cite{yang2025swesmith} & Apr, 2025 & 50,000 & 128 & Synthetic & Semi-manual  \\
      \midrule
      \textbf{\benchmark} (Ours) & April, 2025 & \textbf{1,319 (since 2024)} & \textbf{93} & \textbf{Real} & \textbf{Automatic} \\
    \bottomrule
    \end{tabular}
    }
    \vspace{-1em}
    \label{tab:comparison}
\end{table}

\begin{enumerate}[leftmargin=*]
    \item \textbf{Staleness.} SWE-bench and its derivatives have not been updated since their initial releases, making them static benchmarks. 
	Because LLMs are trained on massive inscrutable corpora, these static datasets are at risk of data contamination, as they could have be been unpurposely included in model training data. 
	This raises concerns about whether newer models are making truly generalizable progress or merely memorizing benchmark content, reducing the benchmarks' effectiveness in distinguishing model capabilities. 

	\item \textbf{Limited repository coverage.} These benchmarks draw from a small set of repositories, limiting diversity in codebases, domains, and programming practices (see Table~\ref{tab:comparison} for details). This narrow scope weakens the generalizability and robustness of evaluations.

	\item \textbf{Heavy reliance on manual effort.} Constructing instances for SWE-bench-like task istances involves substantial human labor: identifying appropriate issue-resolution pairs, locating relevant tests, configuring runnable environments, composing test commands, and validating the full workflow.\footnote{For instance, it take about one year for Multi-SWE-bench \cite{zan2025multi} to create 1,632 benchmark instances with 68 expert annotators. } This process is resource-intensive and creates scalability bottlenecks.
\end{enumerate}

To address these challenges, we introduce \benchmark{}, a live and scalable benchmark built for evaluating LLMs on real-world issue resolution tasks. 
In contrast to recent efforts such as LiveCodeBench~\cite{jain2024livecodebench}, which target algorithmic programming problems, \benchmark{} is the first live-updating benchmark designed for complex, repository-level tasks that demand multi-file reasoning, environment setup, and reproducible execution.
Figure~\ref{fig:overview} illustrates the construction pipeline of \benchmark{}. 
At the core of our framework is \method{}, a fully automated pipeline that eliminates manual bottlenecks by streamlining the entire process—from issue mining to environment packaging.
More specifically, \method{} leverages an agentic and end-to-end workflow to setup the Docker environment by identifying relevant instruction files, selecting base images, installing necessary dependencies, building the project, and validating its test suite.
This automation enables continuous updates, broad repository coverage, and large-scale dataset expansion.
Our current release of \benchmark{} contains 1,319 issue-resolution tasks sourced from real-world GitHub issues created since 2024, spanning 93 repositories.
Compared to existing benchmarks, this represents a significant leap in freshness, diversity, and scale (see Table~\ref{tab:comparison}).

We evaluate three leading agent frameworks (i.e., OpenHands~\cite{openhands}, SWE-Agent~\cite{sweagent}, and Agentless~\cite{xia2024agentless}) in combination with four state-of-the-art LLMs (namely, GPT-4.1, GPT-4o, Claude 3.7 Sonnet, and DeepSeek V3). Consistent with performance rankings reported on SWE-bench Verified,\footnote{\url{https://www.swebench.com/}} we observe that OpenHands, when paired with Claude 3.7 Sonnet, achieves the highest performance on \benchmark{}. However, its overall results are significantly lower compared to those achieved on SWE-bench Verified.
To explore this discrepancy further, we conduct a controlled comparison and find that the same agent-LLM pair consistently performs worse on \benchmark{} than on SWE-bench. This finding suggests that existing models may be overfitting to static benchmarks like SWE-bench, underscoring the importance of developing more dynamic and diverse evaluation settings, such as those provided by \benchmark{}.


Our main contributions are summarized as follows:
\begin{itemize}[leftmargin=*]
\item We introduce \benchmark, a contamination-resistant, reproducible, and continuously updatable benchmark tailored to real-world issue resolution tasks. It reflects the dynamic nature of software development and offers broader repository coverage compared to prior benchmarks.
\item We propose \method, a fully automated pipeline for benchmark construction that seamlessly integrates data curation, environment setup, and test validation into a cohesive and scalable system.
\item Through experimental evaluation, we observe the suboptimal performance of leading agent frameworks on \benchmark, highlighting significant opportunities for improvement on the contamination-free benchmark.
\end{itemize}

\section{Related Work}

\paragraph{Coding Benchmarks.}
Early benchmarks for program synthesis and bug fixing focused on \emph{single‑file, synthetic} tasks such as HumanEval \cite{chen2021evaluating} and MBPP \cite{austin2021program}, which do not reflect the complexity of real repositories.  To move closer to practice, SWE-bench~\cite{swebench} introduced the \emph{issue‐resolving} task, requiring a model to generate a validated patch for a GitHub repositories issue.  Numerous extensions have since appeared—including Multimodal SWE‑bench for JavaScript and UI screenshots~\cite{yang2024swebenchmultimodalaisystems}, Multi-SWE-bench for multiple languages such as Java and Rust~\cite{zan2025multi}.  Despite their impact, all of these datasets are \emph{static}: they are collected once, cover at most a few dozen repositories, and depend on labor‑intensive environment construction.  These yield two limitations. First, models can overfit to the fixed test set, inflating apparent progress.   Second, public tasks may lead to \textit{data contamination}, where benchmark instances leak into pre‑training corpora \cite{zhang2024careful, golchin2023time}.
Recent ``live'' datasets such as LiveCodeBench~\cite{jain2024livecodebench} mitigate contamination by streaming \emph{algorithmic} problems after their release dates, yet they do not address the harder \emph{repository‑level} setting that demands multi‑file reasoning and execution inside a faithful environment. \benchmark is the first open, continuously updating benchmark that fulfills these requirements. 

\paragraph{Coding Agents.}
On top of the above benchmarks, a recent line of work has been working creating autonomous \emph{code agents} that search, edit, and test large codebases.  Representative systems include SWE‑Agent~\cite{yang2024swe}, OpenHands~\cite{openhands}, Agentless~\cite{xia2024agentless}, and training frameworks that synthesize thousands of SWE‑bench‑like instances~\cite{pan2024trainingsoftwareengineeringagents,yang2025swesmith,xie2025swe}.  These agents report remarkable headline numbers, yet their evaluations rely almost exclusively on static offline datasets.  As a consequence, improvements may partially stem from memorisation of leaked solutions or configuration quirks, rather than genuine advances. \benchmark closes this gap by pushing agents to fix \emph{previously unseen, continuously arriving} real‑world bugs under fully reproducible Docker images, it reveals failure modes hidden by stale test suites and provides a trustworthy yard‑stick for code agents and LLMs.

\section{\benchmark}
\label{sec:method}

Targeting the issue resolution task on real-world GitHub repositories, SWE-bench serves as a practical proxy for evaluating the coding capabilities of LLM-based systems. The issue resolving task is defined as follows: given a code repository and an associated issue, an approach (e.g., LLM agent) is required to generate a patch that resolves the issue and passes the test cases (see Appendix~\ref{sec:preliminary} for details). 

While \benchmark{} adopts the same task definition as SWE-bench, it introduces a \emph{novel, fully automated pipeline} that enables scalable and continuously updatable benchmark construction. This automation allows for a larger number of up-to-date instances and broader repository coverage. The initial release of \benchmark consists of \textit{1,319} task instances created between January 2024 and April 2025, spanning \textit{93} real-world repositories. 

\paragraph{Pipeline Overview.} As shown in Figure~\ref{fig:overview}, the construction of \benchmark{} follows a three-stage pipeline. First, starting from popular repositories, we identify GitHub issues that are resolved by a pull request (PR). Next, we apply the proposed \method{}—an agentic approach that automatically sets up an Docker-based execution environment for each candidate instance. Finally, we perform multiple rounds of test execution for each instance to validate whether it consistently exhibits the expected issue-resolving testing behavior, and finalize the valid instances.

Thanks to its fully automated pipeline, \benchmark{} can be maintained with minimal--ideally zero--manual effort. We plan to update \benchmark{} on a monthly basis, continually providing the community with an up-to-date evaluation dataset. This enables contamination-free, rigorous assessment of AI systems’ issue-resolving capabilities in a constantly evolving real-world setting.

\subsection{Raw Issue–PR Crawling}
\label{sec:crawling}

The first phase of the \benchmark pipeline involves collecting real-world issue–pull request (PR) pairs from popular open-source GitHub repositories.

\paragraph{Repository Selection.}
We focus on Python repositories for the initial release of \benchmark, aligning with SWE-bench and other prior benchmarks due to its popularity. The selection process includes three filtering stages:
\textit{(i)} We first queried GitHub API for repositories with over 1,000 stars and Python set as the primary language. This initial query yielded 8,577 repositories as of April 2025.
\textit{(ii)} We then refined this set by requiring each repository to have more than 200 issues and pull requests, over 200 forks, and at least 60\% of its codebase written in Python. This reduced the pool to 3,316 repositories.
\textit{(iii)} Finally, to comply with licensing requirements, we retained only repositories containing a valid open-source license, resulting in a final selection of 2,609 repositories.

\paragraph{Issue–PR Pair Extraction.}
From the selected repositories, we adopt the collection script from SWE-bench to extract issue and its associated PR. Meanwhile, the pull request must modify the repository’s test suite--i.e., a ``test patch'', which will serve as the evaluation targets. We also incorporate improvements from SWE-Fixer~\cite{xie2025swe}, which introduces more robust heuristics to improve the effectiveness of issue–PR pair identification and reduce reliance on the brittle string-matching method. 
To reduce the risk of data leakage, SWE-bench-Live prioritizes recency by including only issues created after January 2024 in our initial release.

\subsection{\method: Automated Execution Environment Setup} 

The ``raw'' issue–PR pairs remain at the textual and plain code level. To support subsequent test-based evaluation, it is required to provide an \textit{execution environment} capable of running tests locally and producing execution feedback. In the context of issue-resolving benchmarks, the execution environment is critical for test-based evaluation.

However, preparing such execution environments is widely recognized as the \textbf{most labour-intensive step} in constructing issue-resolving datasets. In prior work, including SWE-bench \cite{swebench} and SWE-Gym \cite{swegym}, environment setup has been performed entirely by humans. For example, SWE-Gym reports that building execution environments required over 200 hours of manual effort, underscoring a significant scalability bottleneck. Notably, even repository-level environments are insufficient: different commits within the same repository may depend on different libraries or configurations, necessitating environment construction at the \textit{snapshot} level. SWE-bench partially mitigates this by building environments per version tag, but the granularity remains coarse and relies on manual labor.

To address this bottleneck, we introduce an agent-based framework \method, which automatically creates a fully functional execution environment for each issue instance. For any given \textit{repository snapshot}, \method produces a Docker container that installs all required dependencies, builds the project, and validates its test suite. This containerized instance serves as the foundation for running and evaluating model-generated patches.

\paragraph{Repository Snapshots and Environment Definition.}  
A repository snapshot corresponds to the codebase at the base commit associated with an issue. The goal is to recreate an environment faithful to that moment in time. We define a valid execution environment as a Docker container where \textit{(i)} the codebase is correctly installed from source, and \textit{(ii)} the repository's test suite passes with zero or tolerable failures. This environment is essential for test-based evaluation, providing the ground truth mechanism to verify whether the issue has been resolved.

\method{} follows an LLM-driven, agentic workflow \cite{zhang2024large, zhang2025api} inspired by how human developers set up unfamiliar projects, as shown in Figure~\ref{fig:overview}. The process proceeds in five steps:
\begin{itemize}[leftmargin=*]
    \item \textbf{Relevant Files Identification.} The first step is to identify relevant files in the repository--such as CI/CD pipelines and README files that are likely to contain useful information for setting up the environment (a detailed list is provided in the Appendix~\ref{appdix:launchprompt}).
    
    \item \textbf{Base Image Selection.} Given the full content of the relevant files, this step is to select a suitable base Docker image based on the information provided in the repository. This involves correctly identifying the programming language and SDK version used in the repository (e.g., \texttt{python:3.11}). A container is instantiated from the chosen image, and a persistent bash session is launched.
    
    \item \textbf{Interactive Environment Setup.} The setup process is carried out by an agent whose goal is to successfully execute and pass all test cases in the repository’s test suite within the container. The agent interacts with the bash session by issuing commands and receiving feedback such as exit codes and outputs. It follows the ReAct design \cite{yao2023react}, iterating over \textit{Thought} → \textit{Action} → \textit{Observation} \cite{zhang2024ufo, zhang2025ufo2}, mimicking a developer's reasoning and trial process. The agent can also search the web or query the issue tracker for troubleshooting.
    
    \item \textbf{Verification.} Once the setup agent determines that the environment has reached a satisfactory state or a step limit is reached, control is transferred to a verifying agent. The agent attempts to generate the appropriate test command and execute it. The execution results are evaluated with the agent to check if all test cases passed. If test failures occur, the results are fed back to the setup agent for further refinement. If all tests pass, the environment is considered valid.
    
    \item \textbf{Finalization.} Upon successful validation, the container is committed as a Docker image, producing a instance-level execution environment for reuse.
\end{itemize}

\paragraph{Challenges of Version Incompatibility.}  
A major challenge when setting up out-of-date repositories is the ``dependency version drift'' issue. When dependencies are not pinned to specific versions, tools like \texttt{pip} by default will resolve to the latest package versions, which often introduce backward-incompatible issues and make the environment setup fail.  To address this, we implement an \textit{time-machine} mechanism by forcing the package installation tool to only look at valid versions released no later than the current base commit timestamp. Specifically, we modified the \texttt{pip} default index server to a proxy which fetches those valid package versions. This simple but effective strategy prevents the ``future'' version incompatibilities and significantly improves setup success rates.

We will open-source \method to benefit the community. While designed for automated benchmark construction, \method can also assist developers in quickly setting up environments for unfamiliar codebases. Its ability to replicate historical setups and automatically resolve environment dependencies positions it as a practical tool with broader applicability beyond benchmarking.

\subsection{Validating Task Instances}

To ensure the quality of the benchmark, each task instance is validated to confirm that the associated PR effectively resolves the issue it is intended to fix. The validation is based on analyzing changes in the test suite results before and after applying the PR's patch.
Specifically, we focuses on identifying two key behaviors in the test outcomes:
\begin{itemize}[leftmargin=*]
    \item \texttt{FAIL\_TO\_PASS} transitions: Tests that were initially failing (\texttt{FAILED} or \texttt{ERROR}) and later passing (\texttt{PASSED}) after the patch is applied. These yield that the patch addresses the issue effectively.
    \item \texttt{PASS\_TO\_PASS} transitions: Tests that were both passing before and after the patch is applied. These transitions demonstrate that the patch does not break unrelated functionality.
\end{itemize}
To identify these transitions, the test results (as logs) are collected both before and after applying the PR's patch. By comparing individual test outcomes between the two runs, we determine how the patch affected specific tests. We designed framework-specific (e.g., tox, pytest) parsers to interpret test outputs reliably, as different testing tools may produce logs in various formats.
For a task instance to be included in the benchmark, it must exhibit at least one \texttt{FAIL\_TO\_PASS} transition. Instances lacking such a transition are excluded because they do not demonstrate effective bug resolution. Additionally, to ensure reproducibility and avoid issues caused by test flakiness, the validation process is repeated multiple times. Only instances with consistent results across all runs are retained.
This approach ensures that all task instances are grounded in evidence of real-world bug fixes and preserves stable behaviors, resulting in a robust  benchmark for evaluating automated bug-fixing solutions.

\subsection{\benchmark Statistics}

\begin{figure}[t]
    \centering
    \includegraphics[width=0.93\linewidth]{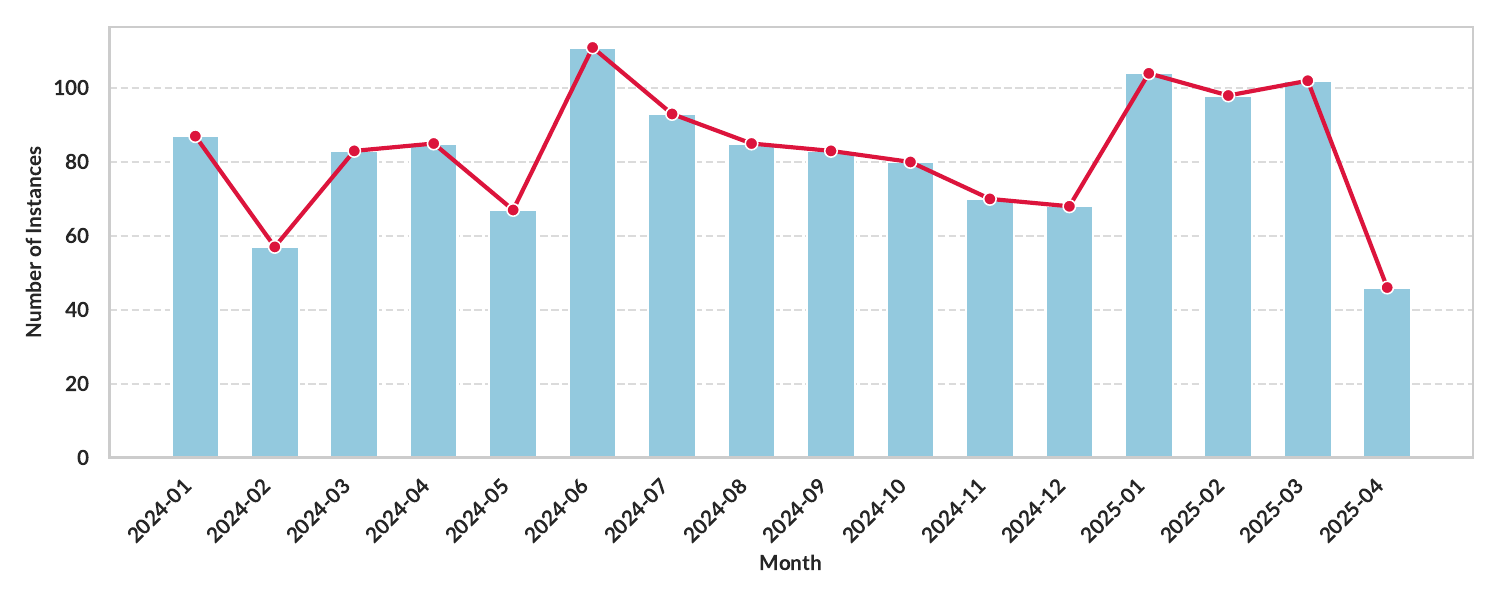}
    \vspace{-1em}
    \caption{Temporal distribution of issue creation times in \benchmark.}
    \vspace{-1.75em}
    \label{fig:created_at}
\end{figure}

The initial release of the \benchmark dataset consists of 1,319 task instances collected from real-world issues and pull requests across 93 open-source Python repositories. To ensure freshness and reduce the risk of data contamination from pretraining, we restrict the dataset to issues created between January 1, 2024, and April 20, 2025. As shown in Figure~\ref{fig:created_at}, the temporal distribution is generally uniform, indicating consistent coverage of issues over time. We plan to update the dataset on a monthly basis to reflect the evolving software landscape and continuously provide new  instances.

\begin{figure}[t]
    \centering
    \begin{minipage}[c]{0.48\textwidth}
        \centering
        \includegraphics[width=\linewidth]{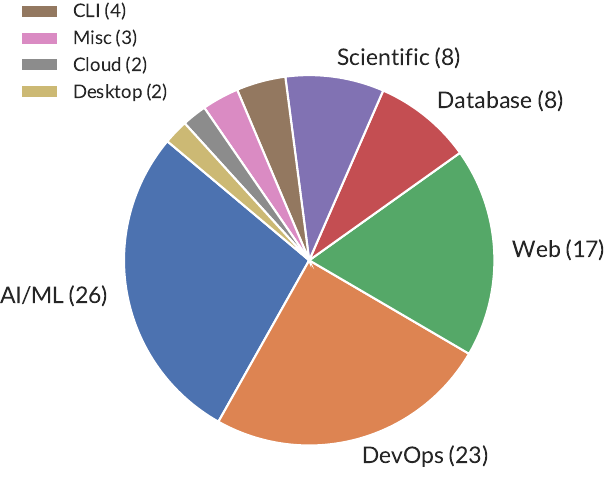}
        \caption{Repository classifications.}
        \label{fig:repo-class}
    \end{minipage}
    \hfill
    \begin{minipage}[c]{0.5\textwidth}
        \centering
        \captionof{table}{Statistics of \benchmark}
        \begin{tabular}{c|c|rr}
        \toprule
          \textbf{Level} & \textbf{\#Item} & \textbf{Average} & \textbf{Median} \\
          \midrule
          \multirow{3}{*}{\rotatebox[origin=c]{90}{Repo}}
            & Repositories & \multicolumn{2}{c}{93} \\
            & LoC\textsuperscript{*} & 85k & 52k \\
            & Files\textsuperscript{*} & 423 & 222 \\
          \midrule
          \multirow{6}{*}{\rotatebox[origin=c]{90}{Instance}}
            & Instances & \multicolumn{2}{c}{1319} \\
            & Files\textsuperscript{\dag} & 3.3 & 2 \\
            & Hunks\textsuperscript{\dag} & 9.0 & 3 \\
            & Lines\textsuperscript{\dag} & 102.6 & 24 \\
            & F2P test cases & 5.4 & 1 \\
            & P2P test cases & 2953.4 & 1865 \\
        \bottomrule
        \multicolumn{4}{l}{\footnotesize{\textsuperscript{*}\textit{Only count Python code.}} \textsuperscript{\dag}\textit{Stats of gold patch.}} \\
        \end{tabular}
        \label{tab:stats}
    \end{minipage}
\vspace{-1em}
\end{figure}

Table~\ref{tab:stats} summarizes key statistics at both the repository and instance levels. At the repository level, projects vary in size, with an average of 85k lines of Python code and 423 files. At the instance level, we report metrics of the gold patches—including the number of edited files, hunks, and lines—as heuristic indicators of task complexity. These statistics suggest that \benchmark tasks reflect realistic, non-trivial bug fixes that challenge code understanding, reasoning, and manipulation capabilities of LLMs. Additionally, we record the number of test cases that transition from failure to pass (\texttt{F2P}) and those that consistently pass (\texttt{P2P}), which form the basis of test-based evaluation.

\paragraph{Repository Diversity.}  
To ensure broad applicability, \benchmark includes repositories from diverse application domains. As shown in Figure~\ref{fig:repo-class}, we manually categorized each repository based on its primary functionality—such as AI/ML, DevOps, Web development, and others. This diversity helps evaluate LLMs across varied software stacks and bug types, enhancing the benchmark’s representativeness of real-world usage scenarios.

\paragraph{Lite Subset.}  
To support lightweight experimentation, we construct a lite subset of \benchmark by sampling 50 instances per month from issues created between October 2024 and March 2025. This results in a compact set of 300 instances that balances recency, diversity, and evaluation efficiency.

\paragraph{Comparison with Existing Benchmarks.}  
Table~\ref{tab:comparison} compares \benchmark with several existing issue-resolution benchmarks. Unlike SWE-bench and its variants, which require extensive manual curation and cover a limited set of repositories, \benchmark is the first to offer an \textbf{automatically} constructed, continuously updatable benchmark. It covers a broader set of repositories (93 in total), while preserving the use of real issues and test-based evaluation. Compared to synthetic datasets like SWE-smith, which may not fully capture the complexity of human-written code and bugs, \benchmark maintains fidelity to real-world development workflows. Its unique combination of automation, realism, and diversity fills a critical gap  of the LLM evaluation for software engineering.

\section{Experiments}
\label{sec:experiments}
\subsection{Setups}

\paragraph{Agents and Model Selection.} 
To evaluate the effectiveness of our proposed \benchmark, we conduct experiments using three representative agent frameworks. These include the general-purpose coding agent \textbf{OpenHands}~\cite{openhands} (paired with CodeAct), as well as two agents specifically designed for issue-resolving tasks: \textbf{SWE-Agent}~\cite{sweagent} and \textbf{Agentless}~\cite{xia2024agentless}. For OpenHands, we set a maximum of 60 iterations per instance. For SWE-Agent, we limit the number of LLM calls to 100 per instance to maintain computational efficiency. For Agentless, we largely follow the original pipeline, which consists of two main stages: issue localization and patch generation. However, we omit the reranking stage based on regression testing, as supporting this step on SWE-bench-Live would require substantial infrastructure adaptation and is beyond the scope of this study. Consequently, both the localization and repair stages in our Agentless evaluation produce a single sample without reranking. We test these agents using four recent state-of-the-art LLMs, covering both proprietary and open-source models: GPT-4o~\cite{openai_gpt4o} (gpt-4o-2024-11-20), GPT-4.1~\cite{openai_gpt41} (gpt-4.1-2025-04-14), Claude 3.7 Sonnet~\cite{claude_37} (claude-3-7-sonnet-20250219), and DeepSeek V3~\cite{deepseek_v3} (DeepSeek-V3-0324).

\paragraph{Evaluation Metrics.} 
Following the evaluation protocol of SWE-bench~\cite{swebench}, we adopt the \textbf{Resolved Rate (\%)} as our primary metric. This measures the proportion of issues successfully resolved by the agent across all task instances. We also report the \textbf{Patch Apply Rate (\%)}, which indicates the percentage of generated patches that are syntactically correct and can be successfully applied to the codebase without errors. Additionally, we measure the \textbf{Localization Success Rate (\%)} at the file level. This reflects whether the set of files modified by the generated patch matches the gold patch.

\subsection{Performance on \benchmark}
We report the performance of all agent–model combinations on the Lite subset of \benchmark{} in Table~\ref{tab:perf_lite}. 
Meanwhile, Table~\ref{tab:full} presents the results of the top three combinations selected based on Lite performance, evaluated on the full version of \benchmark{}.

We observe that the same methods achieve substantially higher scores on SWE-bench compared to their performance on SWE-bench-Live, despite both benchmarks targeting the same issue-resolving task with identical settings. For example, recent state-of-the-art agents and models report a resolved rate exceeding 60\% on the SWE-bench Verified subset\footnote{\url{https://www.swebench.com/}}. In contrast, the highest resolved rate on SWE-bench-Live is only 19.25\%. 
Considering that the experimental setups on the SWE-bench leaderboard often involve dramatically high rollout numbers or iteration efforts, we specifically re-ran the best performing combination, OpenHands with Claude 3.7 Sonnet, on the SWE-bench verified subset using the exact same setups as in our experiments.
The resulting resolved rate reached 43.20\%, more than twice the score achieved on SWE-bench-Live. This is a particularly interesting phenomenon, as it highlights the challenges of constructing a benchmark that can objectively measure an AI system’s ability to resolve arbitrary and previously unseen issues. It also raises concerns about potential overfitting to SWE-bench. Similar phenomena are also observed in other existing issue-resolving datasets: the best-performing method in Multi-SWE-bench achieves a resolved rate of only 19.32\%, while the highest score reported in OmniGIRL is as low as 8.6\%.

\begin{table}
\centering
\caption{Performance on \benchmark (Lite subset).}
\label{tab:perf_lite}
\begin{tabular}{l|rrr}
\toprule
\textbf{Models} & \textbf{Resolved (\%)} & \textbf{Apply (\%)} & \textbf{Loc. Suc. (\%)} \\
\midrule
\rowcolor{gray!20}
\multicolumn{4}{c}{\textbf{OpenHands}} \\
GPT-4o & 7.00 & 72.00 & 30.33 \\
GPT-4.1 & 11.33 & 59.33 & 28.67 \\
Claude 3.7 Sonnet & 17.67 & 84.00 & 48.00 \\
DeepSeek V3 & 13.00 & 81.00 & 38.33 \\
\midrule
\rowcolor{gray!20}
\multicolumn{4}{c}{\textbf{SWE-agent}} \\
GPT-4o & 10.00 & 93.33 & 40.33 \\
GPT-4.1 & 16.33 & 95.00 & 47.33 \\
Claude 3.7 Sonnet & 17.67 & 84.67 & 46.33 \\
DeepSeek V3 & 15.33 & 92.00 & 44.00 \\
\midrule
\rowcolor{gray!20}
\multicolumn{4}{c}{\textbf{Agentless}} \\
GPT-4o & 11.67 & 91.67 & 37.67 \\
GPT-4.1 & 12.00 & 84.33 & 39.00 \\
Claude 3.7 Sonnet & 11.33 & 68.00 & 30.00 \\
DeepSeek V3 & 13.33 & 83.33 & 40.67 \\
\bottomrule

\end{tabular}
\end{table}

\begin{table}[t]
    \centering
    \caption{Performance of top-3 performing Agent + Model combinations on SWE-bench-Live.}
    \begin{tabular}{l|c|rrr}
    \toprule
    \textbf{Agent / Model} & \textbf{Subset} & \textbf{Resolved (\%)} & \textbf{Apply (\%)} & \textbf{Loc. Suc. (\%)} \\
    \midrule
    \multirow{2}{*}{OpenHands / Claude 3.7 Sonnet} & Lite 
    & 17.67 & 84.00 & 48.00 \\ 
  & Full & 19.25 & 85.89 & 48.29 \\
\midrule
\multirow{2}{*}{SWE-agent / GPT-4.1} & Lite
    & 16.33 & 95.00 & 47.33 \\
  & Full  & 18.57 & 94.54 & 49.50 \\
\midrule
\multirow{2}{*}{SWE-agent / Claude 3.7 Sonnet} & Lite
    & 17.67 & 84.67 & 46.33 \\
  & Full & 17.13 & 89.15 & 45.86 \\
    \bottomrule
    \end{tabular}
    
    \vspace{-1em}
    \label{tab:full}
\end{table}

To investigate this, we further categorize the instances in \benchmark{} based on their repository origin. Specifically, 216 instances are derived from 8 repositories that were originally included in SWE-bench, which we refer to as \textit{From SWE-bench Repos}. The remaining 1,103 instances are sourced from repositories not previously used in SWE-bench and are denoted as \textit{From Non-SWE-bench Repos}. As shown in Table~\ref{tab:swe_vs_non_swe}, although the Non-SWE-bench repositories are generally simpler with fewer files and lower code volume, the best-performing agent–model pair achieves a higher resolved rate of 22.96\% on SWE-bench Instances, compared to only 18.89\% on the Non-SWE-bench ones. This reinforces the hypothesis that existing agents may be overfit or implicitly optimized for the SWE-bench repositories, further motivating the need for continuously updated, contamination-resistant benchmarks like SWE-bench-Live.

\begin{table}[h]
    \centering
    \vspace{-1em}
    \caption{SWE-bench vs. Non-SWE-bench.}
    \begin{tabular}{c|ccc}
    \toprule
      Instances & Avg. repo files & Avg. repo loc & 
      Resolved (\%) \\
    \midrule
      From SWE-bench Repos & 744 & 223k & 22.96 \\
      From Non-SWE-bench Repos & 383 & 68k & 18.89  \\
    \bottomrule
    \end{tabular}
    \vspace{-1em}
    \label{tab:swe_vs_non_swe}
\end{table}

\subsection{Performance vs. Creation Date.}
\begin{figure}[!t]
    \centering
    \begin{minipage}[c]{0.58\textwidth}
        \centering
        \includegraphics[width=\linewidth]{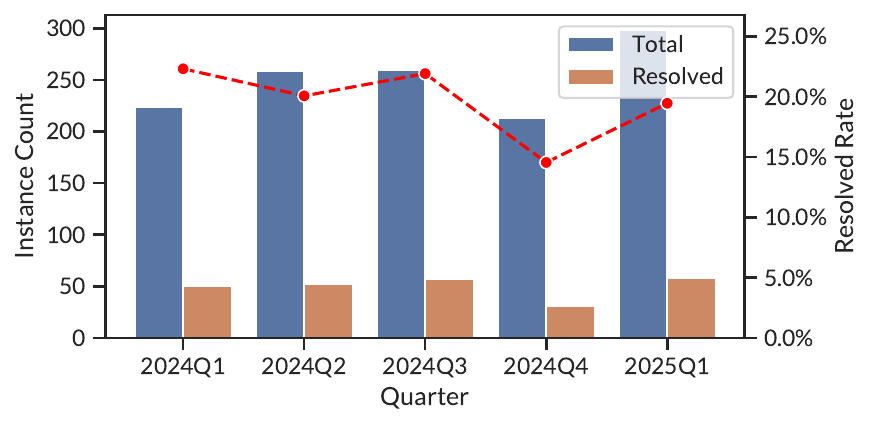}
        \caption{Resolved rate in relation to the creation date of instances. (OpenHands / Claude 3.7 Sonnet on full set)}
        \label{fig:perf_vs_time}
    \end{minipage}
    \hfill
    \begin{minipage}[c]{0.4\textwidth}
        \centering
        \includegraphics[width=\linewidth]{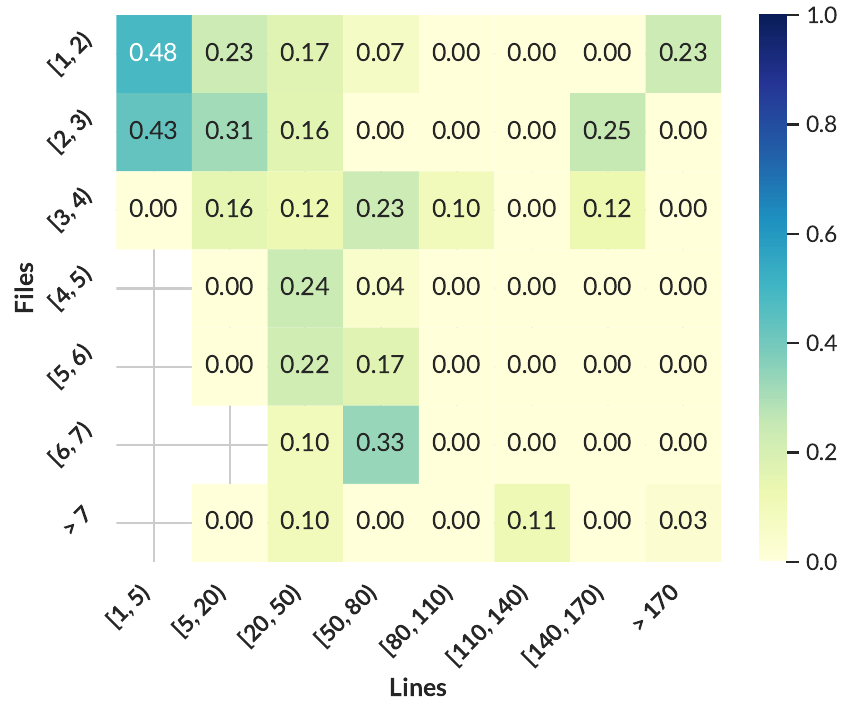}
        \caption{Resolved rate in relation to the difficulty of instances. (OpenHands / Claude 3.7 Sonnet on full set)}
        \label{fig:perf_vs_difficulty}
    \end{minipage}
\end{figure}

To investigate whether the recency of an issue affects its difficulty, we analyze the resolved rate across different creation periods. As shown in Figure~\ref{fig:perf_vs_time}, SWE-bench-Live includes a balanced distribution of instances across quarters from 2024Q1 to 2025Q1. The resolved rate, based on OpenHands with Claude 3.7 Sonnet on the full benchmark, remains relatively stable over time, fluctuating only modestly across quarters.

While there is a slight dip in resolved rate during 2024Q4, followed by a recovery in 2025Q1, the trend does not indicate a clear correlation between task recency and success rate. This suggests that newer issues are not inherently harder for current agents to solve, and that SWE-bench-Live maintains a consistent level of challenge across time. These results reinforce the benchmark's ability to deliver a steady and reliable evaluation signal, even as it continuously evolves with newly introduced instances.

\subsection{Performance vs. Difficulty}
\label{sec:perform_vs_diff}
We approximate the difficulty of a bug–fixing instance along two complementary axes.  
\emph{Patch difficulty} is captured by the scope of the gold fix—the number of files it touches and the total lines modified—while \emph{repository difficulty} is approximated by the overall size of the project in files and lines of code (LOC).

\textbf{Patch difficulty.}  
Figure~\ref{fig:perf_vs_difficulty} visualises resolved rate as a heat-map over patch scope.  Success is high when the fix is local: a single-file patch that changes fewer than five lines is solved almost one time in two (48\%).  Performance degrades quickly as either dimension grows.  Once the patch edits three or more files, or spans more than one hundred lines, the success rate falls below ten per-cent; patches that touch seven or more files are never solved.  The sharp drop beyond the  \emph{one-file / few-lines} corner highlights a key limitation of current agents: they struggle to coordinate coherent edits across multiple files or to reason about large, intra-file changes.

\textbf{Repository difficulty.}  
Figure~\ref{fig:perf_with_repo} in Appendix~\ref{appendix:perform_vs_repo} plots resolved rate for every repository against its size (Python files on the x-axis, LOC on the y-axis).  Bubble area reflects the number of instances drawn from each project, and red outlines mark the original SWE-bench repositories.  A clear negative trend emerges: repositories with fewer than one hundred files and under twenty-thousand LOC often yield success rates above twenty per-cent, whereas projects exceeding five-hundred files rarely exceed five per-cent.  Nevertheless, notable variance remains—some small-to-mid-size projects are still hard to fix, likely due to atypical build systems or complex domain logic—emphasising that size is an informative but imperfect proxy for difficulty.

Together, the two figures show that difficulty increases along both local (patch) and global (repository) dimensions, and that current code agents falter once fixes spill beyond a handful of lines or involve cross-file reasoning.  Because SWE-bench-Live spans the full spectrum of these difficulty factors—while continuously adding fresh, unseen instances—it provides a stringent and up-to-date testbed for future advances in large-scale program repair.

\section{Conclusion}
We present \benchmark{}, the first continuously updating benchmark for evaluating large language models on real-world issue resolution tasks at the repository level for fresh issue fixing. By addressing key limitations of prior benchmarks such as dataset staleness, limited repository diversity, and manual curation cost, \benchmark{} provides a scalable, contamination resistant, and fully automated evaluation framework. At its core is \method{}, an agent based pipeline that builds reproducible Docker environments and validates issue and pull request pairs through test execution, removing the need for manual intervention. Our empirical results across multiple agent and model combinations show that \benchmark{} presents significantly greater challenges than static datasets. The low resolution rates, especially on multi file patches and large codebases, highlight the limitations of current systems and the importance of live benchmarks in measuring true model generalization.

\bibliographystyle{plain}
\bibliography{references}

\begin{thebibliography}{10}

\bibitem{claude_37}
{Anthropic}.
\newblock Claude 3.7 sonnet and claude code, 2025.
\newblock Accessed: 2025-05-14.

\bibitem{austin2021program}
Jacob Austin, Augustus Odena, Maxwell Nye, Maarten Bosma, Henryk Michalewski, David Dohan, Ellen Jiang, Carrie Cai, Michael Terry, Quoc Le, et~al.
\newblock Program synthesis with large language models.
\newblock {\em arXiv preprint arXiv:2108.07732}, 2021.

\bibitem{chen2021evaluating}
Mark Chen, Jerry Tworek, Heewoo Jun, Qiming Yuan, Henrique Ponde De~Oliveira Pinto, Jared Kaplan, Harri Edwards, Yuri Burda, Nicholas Joseph, Greg Brockman, et~al.
\newblock Evaluating large language models trained on code.
\newblock {\em arXiv preprint arXiv:2107.03374}, 2021.

\bibitem{cursor2025}
{Cursor}.
\newblock Cursor – the ai-powered code editor, 2025.
\newblock Accessed: 2025-05-14.

\bibitem{dakhel2023github}
Arghavan~Moradi Dakhel, Vahid Majdinasab, Amin Nikanjam, Foutse Khomh, Michel~C Desmarais, and Zhen Ming~Jack Jiang.
\newblock Github copilot ai pair programmer: Asset or liability?
\newblock {\em Journal of Systems and Software}, 203:111734, 2023.

\bibitem{deepseek_v3}
DeepSeek-AI.
\newblock Deepseek-v3 technical report, 2025.

\bibitem{fan2023large}
Angela Fan, Beliz Gokkaya, Mark Harman, Mitya Lyubarskiy, Shubho Sengupta, Shin Yoo, and Jie~M Zhang.
\newblock Large language models for software engineering: Survey and open problems.
\newblock In {\em 2023 IEEE/ACM International Conference on Software Engineering: Future of Software Engineering (ICSE-FoSE)}, pages 31--53. IEEE, 2023.

\bibitem{golchin2023time}
Shahriar Golchin and Mihai Surdeanu.
\newblock Time travel in llms: Tracing data contamination in large language models.
\newblock {\em arXiv preprint arXiv:2308.08493}, 2023.

\bibitem{jain2024livecodebench}
Naman Jain, King Han, Alex Gu, Wen-Ding Li, Fanjia Yan, Tianjun Zhang, Sida Wang, Armando Solar-Lezama, Koushik Sen, and Ion Stoica.
\newblock Livecodebench: Holistic and contamination free evaluation of large language models for code.
\newblock {\em arXiv preprint arXiv:2403.07974}, 2024.

\bibitem{swebench}
Carlos~E Jimenez, John Yang, Alexander Wettig, Shunyu Yao, Kexin Pei, Ofir Press, and Karthik~R Narasimhan.
\newblock Swe-bench: Can language models resolve real-world github issues?
\newblock In {\em The Twelfth International Conference on Learning Representations}.

\bibitem{openai_gpt4o}
{OpenAI}.
\newblock Hello gpt-4o, 2025.
\newblock Accessed: 2025-05-14.

\bibitem{openai_gpt41}
{OpenAI}.
\newblock Introducing gpt-4.1 in the api, 2025.
\newblock Accessed: 2025-05-14.

\bibitem{swebench-verified}
{OpenAI}.
\newblock Introducing swe-bench verified, 2025.
\newblock Accessed: 2025-05-05.

\bibitem{swegym}
Jiayi Pan, Xingyao Wang, Graham Neubig, Navdeep Jaitly, Heng Ji, Alane Suhr, and Yizhe Zhang.
\newblock Training software engineering agents and verifiers with swe-gym.
\newblock {\em arXiv preprint arXiv:2412.21139}, 2024.

\bibitem{pan2024trainingsoftwareengineeringagents}
Jiayi Pan, Xingyao Wang, Graham Neubig, Navdeep Jaitly, Heng Ji, Alane Suhr, and Yizhe Zhang.
\newblock Training software engineering agents and verifiers with swe-gym, 2024.

\bibitem{openhands}
Xingyao Wang, Boxuan Li, Yufan Song, Frank~F. Xu, Xiangru Tang, Mingchen Zhuge, Jiayi Pan, Yueqi Song, Bowen Li, Jaskirat Singh, Hoang~H. Tran, Fuqiang Li, Ren Ma, Mingzhang Zheng, Bill Qian, Yanjun Shao, Niklas Muennighoff, Yizhe Zhang, Binyuan Hui, Junyang Lin, Robert Brennan, Hao Peng, Heng Ji, and Graham Neubig.
\newblock {OpenHands: An Open Platform for AI Software Developers as Generalist Agents}, 2024.

\bibitem{xia2024agentless}
Chunqiu~Steven Xia, Yinlin Deng, Soren Dunn, and Lingming Zhang.
\newblock Agentless: Demystifying llm-based software engineering agents.
\newblock {\em arXiv preprint arXiv:2407.01489}, 2024.

\bibitem{xie2025swe}
Chengxing Xie, Bowen Li, Chang Gao, He~Du, Wai Lam, Difan Zou, and Kai Chen.
\newblock Swe-fixer: Training open-source llms for effective and efficient github issue resolution.
\newblock {\em arXiv preprint arXiv:2501.05040}, 2025.

\bibitem{xuopenrca}
Junjielong Xu, Qinan Zhang, Zhiqing Zhong, Shilin He, Chaoyun Zhang, Qingwei Lin, Dan Pei, Pinjia He, Dongmei Zhang, and Qi~Zhang.
\newblock Openrca: Can large language models locate the root cause of software failures?
\newblock In {\em The Thirteenth International Conference on Learning Representations}.

\bibitem{sweagent}
John Yang, Carlos Jimenez, Alexander Wettig, Kilian Lieret, Shunyu Yao, Karthik Narasimhan, and Ofir Press.
\newblock Swe-agent: Agent-computer interfaces enable automated software engineering.
\newblock {\em Advances in Neural Information Processing Systems}, 37:50528--50652, 2024.

\bibitem{yang2024swe}
John Yang, Carlos Jimenez, Alexander Wettig, Kilian Lieret, Shunyu Yao, Karthik Narasimhan, and Ofir Press.
\newblock Swe-agent: Agent-computer interfaces enable automated software engineering.
\newblock {\em Advances in Neural Information Processing Systems}, 37:50528--50652, 2024.

\bibitem{yang2024swebenchmultimodalaisystems}
John Yang, Carlos~E. Jimenez, Alex~L. Zhang, Kilian Lieret, Joyce Yang, Xindi Wu, Ori Press, Niklas Muennighoff, Gabriel Synnaeve, Karthik~R. Narasimhan, Diyi Yang, Sida~I. Wang, and Ofir Press.
\newblock Swe-bench multimodal: Do ai systems generalize to visual software domains?, 2024.

\bibitem{yang2025swesmith}
John Yang, Kilian Leret, Carlos~E. Jimenez, Alexander Wettig, Kabir Khandpur, Yanzhe Zhang, Binyuan Hui, Ofir Press, Ludwig Schmidt, and Diyi Yang.
\newblock Swe-smith: Scaling data for software engineering agents, 2025.

\bibitem{yao2023react}
Shunyu Yao, Jeffrey Zhao, Dian Yu, Nan Du, Izhak Shafran, Karthik Narasimhan, and Yuan Cao.
\newblock React: Synergizing reasoning and acting in language models.
\newblock In {\em International Conference on Learning Representations (ICLR)}, 2023.

\bibitem{zan2025multi}
Daoguang Zan, Zhirong Huang, Wei Liu, Hanwu Chen, Linhao Zhang, Shulin Xin, Lu~Chen, Qi~Liu, Xiaojian Zhong, Aoyan Li, et~al.
\newblock Multi-swe-bench: A multilingual benchmark for issue resolving.
\newblock {\em arXiv preprint arXiv:2504.02605}, 2025.

\bibitem{zhang2025api}
Chaoyun Zhang, Shilin He, Liqun Li, Si~Qin, Yu~Kang, Qingwei Lin, and Dongmei Zhang.
\newblock Api agents vs. gui agents: Divergence and convergence.
\newblock {\em arXiv preprint arXiv:2503.11069}, 2025.

\bibitem{zhang2024large}
Chaoyun Zhang, Shilin He, Jiaxu Qian, Bowen Li, Liqun Li, Si~Qin, Yu~Kang, Minghua Ma, Guyue Liu, Qingwei Lin, et~al.
\newblock Large language model-brained gui agents: A survey.
\newblock {\em arXiv preprint arXiv:2411.18279}, 2024.

\bibitem{zhang2025ufo2}
Chaoyun Zhang, He~Huang, Chiming Ni, Jian Mu, Si~Qin, Shilin He, Lu~Wang, Fangkai Yang, Pu~Zhao, Chao Du, et~al.
\newblock Ufo2: The desktop agentos.
\newblock {\em arXiv preprint arXiv:2504.14603}, 2025.

\bibitem{zhang2024ufo}
Chaoyun Zhang, Liqun Li, Shilin He, Xu~Zhang, Bo~Qiao, Si~Qin, Minghua Ma, Yu~Kang, Qingwei Lin, Saravan Rajmohan, et~al.
\newblock Ufo: A ui-focused agent for windows os interaction.
\newblock {\em arXiv preprint arXiv:2402.07939}, 2024.

\bibitem{zhang2024careful}
Hugh Zhang, Jeff Da, Dean Lee, Vaughn Robinson, Catherine Wu, William Song, Tiffany Zhao, Pranav Raja, Charlotte Zhuang, Dylan Slack, et~al.
\newblock A careful examination of large language model performance on grade school arithmetic.
\newblock {\em Advances in Neural Information Processing Systems}, 37:46819--46836, 2024.

\bibitem{zhang2025di}
Linghao Zhang, Junhao Wang, Shilin He, Chaoyun Zhang, Yu~Kang, Bowen Li, Jiaheng Wen, Chengxing Xie, Maoquan Wang, Yufan Huang, et~al.
\newblock Di-bench: Benchmarking large language models on dependency inference with testable repositories at scale.
\newblock {\em arXiv preprint arXiv:2501.13699}, 2025.

\end{thebibliography}

\newpage
\appendix

\section{Full Repositories List}

\resizebox{\textwidth}{!}{
\begin{tabular}{llrrrr}
\toprule
Type & Repository & License & \#Instances & \#Files & LoC \\
\midrule
\multirow{26}{*}{AI/ML} & \href{https://github.com/deepset-ai/haystack}{deepset-ai/haystack} & Apache-2.0 & 64 & 433 & 84.8k \\
& \href{https://github.com/instructlab/instructlab}{instructlab/instructlab} & Apache-2.0 & 52 & 142 & 28.2k \\
& \href{https://github.com/keras-team/keras}{keras-team/keras} & Apache-2.0 & 48 & 900 & 249.7k \\
& \href{https://github.com/kedro-org/kedro}{kedro-org/kedro} & Apache-2.0 & 27 & 179 & 40.4k \\
& \href{https://github.com/pytorch/torchtune}{pytorch/torchtune} & BSD-3-Clause & 14 & 448 & 92.4k \\
& \href{https://github.com/jupyterlab/jupyter-ai}{jupyterlab/jupyter-ai} & BSD-3-Clause & 13 & 81 & 9.0k \\
& \href{https://github.com/run-llama/llama\_deploy}{run-llama/llama\_deploy} & MIT & 12 & 216 & 15.1k \\
& \href{https://github.com/stanfordnlp/dspy}{stanfordnlp/dspy} & MIT & 10 & 222 & 30.6k \\
& \href{https://github.com/projectmesa/mesa}{projectmesa/mesa} & Apache-2.0 & 9 & 109 & 20.3k \\
& \href{https://github.com/huggingface/smolagents}{huggingface/smolagents} & Apache-2.0 & 5 & 65 & 21.4k \\
& \href{https://github.com/theOehrly/Fast-F1}{theOehrly/Fast-F1} & MIT & 4 & 92 & 20.7k \\
& \href{https://github.com/cyclotruc/gitingest}{cyclotruc/gitingest} & MIT & 3 & 39 & 4.7k \\
& \href{https://github.com/modelcontextprotocol/python-sdk}{modelcontextprotocol/python-sdk} & MIT & 2 & 114 & 13.4k \\
& \href{https://github.com/camel-ai/camel}{camel-ai/camel} & Apache-2.0 & 2 & 799 & 130.1k \\
& \href{https://github.com/hiyouga/LLaMA-Factory}{hiyouga/LLaMA-Factory} & Apache-2.0 & 2 & 170 & 31.2k \\
& \href{https://github.com/feast-dev/feast}{feast-dev/feast} & Apache-2.0 & 2 & 673 & 103.3k \\
& \href{https://github.com/openai/openai-agents-python}{openai/openai-agents-python} & MIT & 1 & 212 & 29.8k \\
& \href{https://github.com/huggingface/datasets}{huggingface/datasets} & Apache-2.0 & 1 & 207 & 69.6k \\
& \href{https://github.com/stanford-crfm/helm}{stanford-crfm/helm} & Apache-2.0 & 1 & 891 & 122.1k \\
& \href{https://github.com/freqtrade/freqtrade}{freqtrade/freqtrade} & GPL-3.0 & 1 & 458 & 130.1k \\
& \href{https://github.com/lss233/kirara-ai}{lss233/kirara-ai} & AGPL-3.0 & 1 & 261 & 25.5k \\
& \href{https://github.com/arviz-devs/arviz}{arviz-devs/arviz} & Apache-2.0 & 1 & 259 & 50.6k \\
& \href{https://github.com/qubvel-org/segmentation\_models.pytorch}{qubvel-org/segmentation\_models.pytorch} & MIT & 1 & 130 & 18.6k \\
& \href{https://github.com/scikit-learn-contrib/category\_encoders}{scikit-learn-contrib/category\_encoders} & BSD-3-Clause & 1 & 71 & 12.9k \\
& \href{https://github.com/huggingface/open-r1}{huggingface/open-r1} & Apache-2.0 & 1 & 29 & 4.0k \\
& \href{https://github.com/gptme/gptme}{gptme/gptme} & MIT & 1 & 124 & 23.3k \\
\midrule
\multirow{23}{*}{DevOps} & \href{https://github.com/conan-io/conan}{conan-io/conan} & MIT & 136 & 1056 & 162.5k \\
& \href{https://github.com/pylint-dev/pylint}{pylint-dev/pylint} & GPL-2.0 & 57 & 2301 & 116.8k \\
& \href{https://github.com/sphinx-doc/sphinx}{sphinx-doc/sphinx} & N/A & 39 & 718 & 140.3k \\
& \href{https://github.com/pdm-project/pdm}{pdm-project/pdm} & MIT & 34 & 221 & 32.3k \\
& \href{https://github.com/beeware/briefcase}{beeware/briefcase} & BSD-3-Clause & 24 & 508 & 89.3k \\
& \href{https://github.com/bridgecrewio/checkov}{bridgecrewio/checkov} & Apache-2.0 & 21 & 4551 & 234.9k \\
& \href{https://github.com/joke2k/faker}{joke2k/faker} & MIT & 20 & 754 & 351.4k \\
& \href{https://github.com/python-attrs/attrs}{python-attrs/attrs} & MIT & 10 & 52 & 18.6k \\
& \href{https://github.com/ipython/ipython}{ipython/ipython} & BSD-3-Clause & 10 & 293 & 79.8k \\
& \href{https://github.com/koxudaxi/datamodel-code-generator}{koxudaxi/datamodel-code-generator} & MIT & 10 & 599 & 60.3k \\
& \href{https://github.com/tox-dev/tox}{tox-dev/tox} & MIT & 7 & 225 & 23.8k \\
& \href{https://github.com/dynaconf/dynaconf}{dynaconf/dynaconf} & MIT & 6 & 463 & 55.1k \\
& \href{https://github.com/pypa/twine}{pypa/twine} & Apache-2.0 & 6 & 34 & 6.6k \\
& \href{https://github.com/wemake-services/wemake-python-styleguide}{wemake-services/wemake-python-styleguide} & MIT & 6 & 396 & 52.5k \\
& \href{https://github.com/Delgan/loguru}{Delgan/loguru} & MIT & 6 & 168 & 19.2k \\
& \href{https://github.com/kubernetes-client/python}{kubernetes-client/python} & Apache-2.0 & 3 & 783 & 267.2k \\
& \href{https://github.com/olofk/fusesoc}{olofk/fusesoc} & BSD-2-Clause & 2 & 45 & 8.8k \\
& \href{https://github.com/amoffat/sh}{amoffat/sh} & MIT & 2 & 5 & 7.4k \\
& \href{https://github.com/facebookresearch/hydra}{facebookresearch/hydra} & MIT & 2 & 439 & 41.4k \\
& \href{https://github.com/home-assistant/supervisor}{home-assistant/supervisor} & Apache-2.0 & 2 & 541 & 82.3k \\
& \href{https://github.com/FreeOpcUa/opcua-asyncio}{FreeOpcUa/opcua-asyncio} & LGPL-3.0 & 1 & 168 & 344.4k \\
& \href{https://github.com/pytest-dev/pytest}{pytest-dev/pytest} & MIT & 1 & 260 & 99.5k \\
& \href{https://github.com/iterative/dvc}{iterative/dvc} & Apache-2.0 & 1 & 554 & 85.3k \\
\midrule
\multirow{12}{*}{Web} & \href{https://github.com/reflex-dev/reflex}{reflex-dev/reflex} & Apache-2.0 & 44 & 376 & 89.8k \\
& \href{https://github.com/sissbruecker/linkding}{sissbruecker/linkding} & MIT & 29 & 193 & 26.4k \\
& \href{https://github.com/Kozea/WeasyPrint}{Kozea/WeasyPrint} & BSD-3-Clause & 19 & 144 & 70.0k \\
& \href{https://github.com/python-telegram-bot/python-telegram-bot}{python-telegram-bot/python-telegram-bot} & GPL-3.0 & 16 & 464 & 140.8k \\
& \href{https://github.com/python-babel/babel}{python-babel/babel} & BSD-3-Clause & 11 & 75 & 23.1k \\
& \href{https://github.com/falconry/falcon}{falconry/falcon} & Apache-2.0 & 11 & 262 & 58.5k \\
& \href{https://github.com/aiogram/aiogram}{aiogram/aiogram} & MIT & 11 & 861 & 69.8k \\
& \href{https://github.com/privacyidea/privacyidea}{privacyidea/privacyidea} & AGPL-3.0 & 10 & 483 & 167.5k \\
& \href{https://github.com/urllib3/urllib3}{urllib3/urllib3} & MIT & 10 & 81 & 31.3k \\
& \href{https://github.com/ag2ai/faststream}{ag2ai/faststream} & Apache-2.0 & 6 & 1267 & 85.1k \\
& \href{https://github.com/encode/starlette}{encode/starlette} & BSD-3-Clause & 5 & 66 & 17.2k \\
& \href{https://github.com/scrapinghub/dateparser}{scrapinghub/dateparser} & BSD-3-Clause & 2 & 274 & 67.1k \\
\bottomrule
\end{tabular}
}


\resizebox{\textwidth}{!}{
\begin{tabular}{llrrrr}
\toprule
Type & Repository & License & \#Instances & \#Files & LoC \\
\midrule
\multirow{5}{*}{Web} & \href{https://github.com/pallets/flask}{pallets/flask} & BSD-3-Clause & 2 & 83 & 17.8k \\
& \href{https://github.com/scrapy-plugins/scrapy-splash}{scrapy-plugins/scrapy-splash} & BSD-3-Clause & 1 & 25 & 3.4k \\
& \href{https://github.com/psf/requests}{psf/requests} & Apache-2.0 & 1 & 36 & 11.2k \\
& \href{https://github.com/jpadilla/pyjwt}{jpadilla/pyjwt} & MIT & 1 & 26 & 6.9k \\
& \href{https://github.com/slackapi/bolt-python}{slackapi/bolt-python} & MIT & 1 & 562 & 60.8k \\
\midrule
\multirow{8}{*}{Database} & \href{https://github.com/pydata/xarray}{pydata/xarray} & Apache-2.0 & 29 & 226 & 179.2k \\
& \href{https://github.com/geopandas/geopandas}{geopandas/geopandas} & BSD-3-Clause & 21 & 87 & 47.3k \\
& \href{https://github.com/reata/sqllineage}{reata/sqllineage} & MIT & 18 & 103 & 9.7k \\
& \href{https://github.com/patroni/patroni}{patroni/patroni} & MIT & 17 & 117 & 45.9k \\
& \href{https://github.com/piskvorky/smart\_open}{piskvorky/smart\_open} & MIT & 6 & 64 & 12.4k \\
& \href{https://github.com/wireservice/csvkit}{wireservice/csvkit} & MIT & 3 & 48 & 6.6k \\
& \href{https://github.com/jazzband/tablib}{jazzband/tablib} & MIT & 2 & 32 & 6.6k \\
& \href{https://github.com/Flexget/Flexget}{Flexget/Flexget} & MIT & 2 & 657 & 108.6k \\
\midrule
\multirow{8}{*}{Scientific} & \href{https://github.com/pvlib/pvlib-python}{pvlib/pvlib-python} & BSD-3-Clause & 29 & 178 & 59.9k \\
& \href{https://github.com/python-control/python-control}{python-control/python-control} & BSD-3-Clause & 15 & 155 & 70.7k \\
& \href{https://github.com/mikedh/trimesh}{mikedh/trimesh} & MIT & 14 & 248 & 74.8k \\
& \href{https://github.com/PyPSA/PyPSA}{PyPSA/PyPSA} & MIT & 10 & 129 & 32.3k \\
& \href{https://github.com/shapely/shapely}{shapely/shapely} & BSD-3-Clause & 9 & 158 & 34.0k \\
& \href{https://github.com/pybamm-team/PyBaMM}{pybamm-team/PyBaMM} & BSD-3-Clause & 6 & 581 & 113.4k \\
& \href{https://github.com/beancount/beancount}{beancount/beancount} & GPL-2.0 & 2 & 194 & 48.0k \\
& \href{https://github.com/sympy/sympy}{sympy/sympy} & N/A & 2 & 1574 & 760.5k \\
\midrule
\multirow{4}{*}{CLI} & \href{https://github.com/streamlink/streamlink}{streamlink/streamlink} & BSD-2-Clause & 39 & 510 & 84.0k \\
& \href{https://github.com/beetbox/beets}{beetbox/beets} & MIT & 9 & 193 & 69.3k \\
& \href{https://github.com/yt-dlp/yt-dlp}{yt-dlp/yt-dlp} & Unlicense & 5 & 1177 & 244.6k \\
& \href{https://github.com/jarun/buku}{jarun/buku} & GPL-3.0 & 2 & 27 & 7.1k \\
\midrule
\multirow{3}{*}{Misc} & \href{https://github.com/matplotlib/matplotlib}{matplotlib/matplotlib} & N/A & 85 & 904 & 263.6k \\
& \href{https://github.com/fonttools/fonttools}{fonttools/fonttools} & MIT & 12 & 512 & 192.6k \\
& \href{https://github.com/pytransitions/transitions}{pytransitions/transitions} & MIT & 2 & 39 & 12.4k \\
\midrule
\multirow{2}{*}{Cloud} & \href{https://github.com/aws-cloudformation/cfn-lint}{aws-cloudformation/cfn-lint} & MIT-0 & 102 & 2422 & 160.2k \\
& \href{https://github.com/icloud-photos-downloader/icloud\_photos\_downloader}{icloud-photos-downloader/icloud\_photos\_downloader} & MIT & 4 & 73 & 15.5k \\
\midrule
\multirow{2}{*}{Desktop} & \href{https://github.com/qtile/qtile}{qtile/qtile} & MIT & 6 & 405 & 81.6k \\
& \href{https://github.com/pwr-Solaar/Solaar}{pwr-Solaar/Solaar} & GPL-2.0 & 3 & 94 & 33.7k \\
\bottomrule
\end{tabular}
}

\section{Task Formulation}
\label{sec:preliminary}

\begin{figure}[!h]
    \centering
    \includegraphics[width=0.8\linewidth]{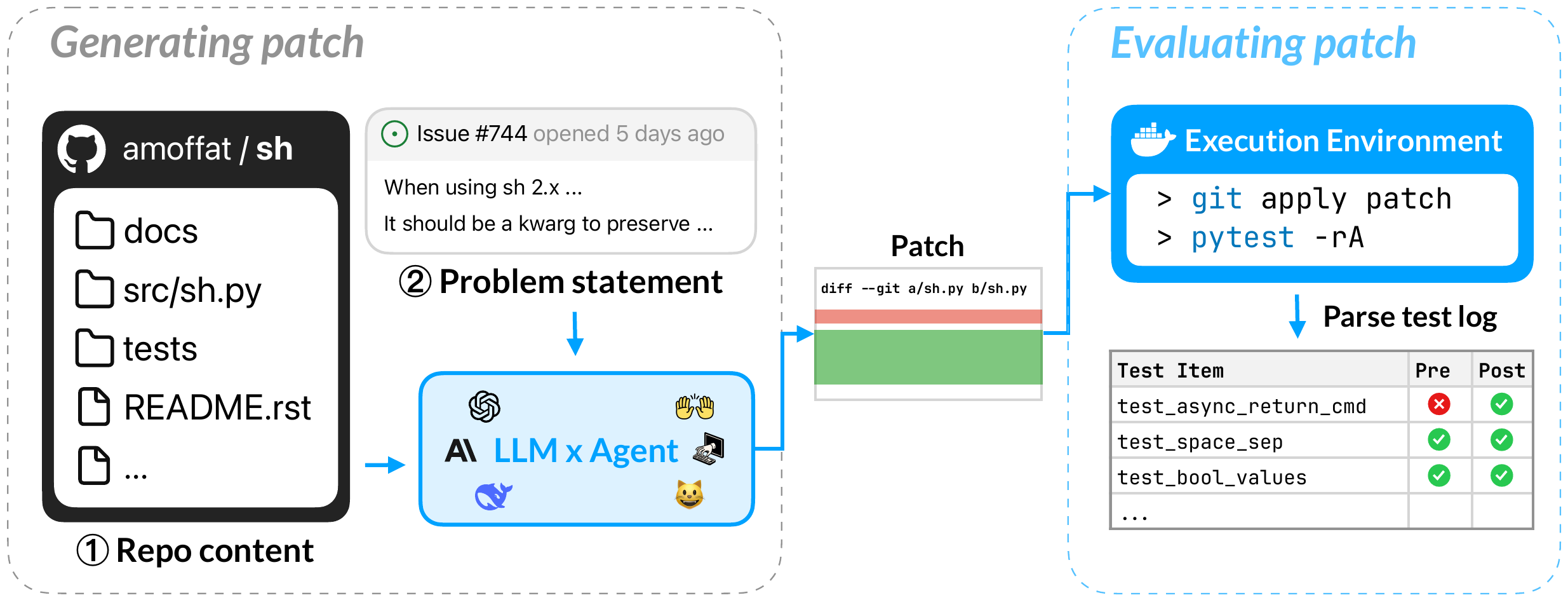}
    \caption{The issue-resolving task requires the model to generate a patch that addresses a given issue, with its correctness evaluated through test execution.}
    \label{fig:task}
\end{figure}

Issue resolving is the task that introduced by SWE-bench~\cite{swebench} for benchmarking AI coding capabilities. In simple terms, it simulates the process of a developer submitting a pull request to address an issue. The formulation of the issue-resolving task is illustrated in Figure~\ref{fig:task}.

\paragraph{Generating Patch.} The task input includes the problem statement of the issue, which is the description written by the issue reporter, as well as a snapshot of the codebase at the time the issue was filed (obtained by resetting to the \texttt{base\_commit}). The model has access to full content of the codebase, after then it is tasked with generating a patch that fixes the given issue, analogous to the file changes submitted in a pull request. In practice, the expected output is in the \texttt{.diff} format.

\paragraph{Evaluating Patch.} Once a patch is proposed by the model, we assess its correctness by applying it to the target codebase and executing the repository’s test suite. The output of the test execution are parsed using a log parser function, which extracts the status of each individual test case. These results are then compared against the expected test case transitions pre-defined for the issue, specifically \texttt{FAIL\_TO\_PASS} and \texttt{PASS\_TO\_PASS}. \texttt{FAIL\_TO\_PASS} refers to test cases that originally failed prior to patch application—typically those introduced in the corresponding pull request—and are expected to pass if the proposed solution is correct. A correct patch should successfully cause these failing tests to pass, without causing regressions in the already passing tests.

\section{Performance vs Repository difficulty}
\label{appendix:perform_vs_repo}
The following Figure~\ref{fig:perf_with_repo} plots resolved rate for every repository against its size (Python files on the x-axis, LOC on the y-axis). For detailed interpreatation of the figure, please see Section \ref{sec:perform_vs_diff}.
\begin{figure}[h!]
    \centering
    \includegraphics[width=\linewidth]{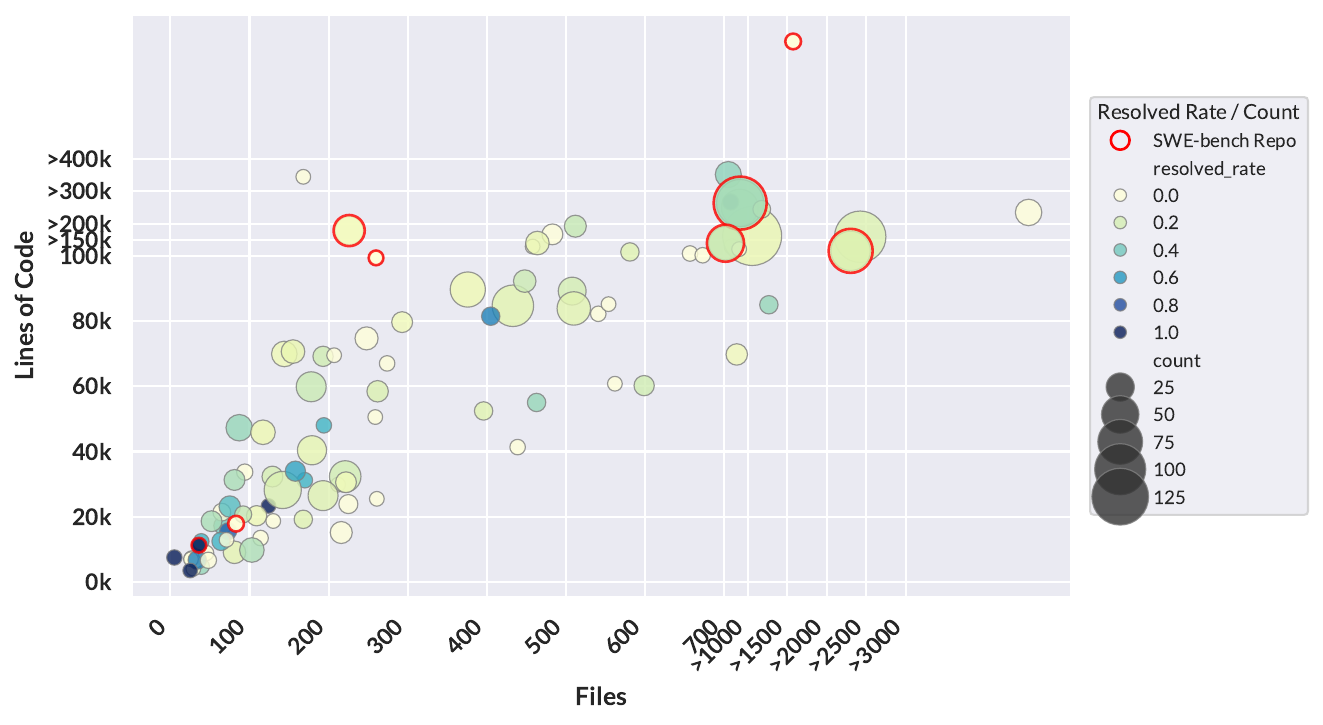}
    \caption{Resolved rate in relation to the number of files and lines of code of a repository. }
    \label{fig:perf_with_repo}
\end{figure}

\section{Dataset Fields}

Table~\ref{tab:fields} provides a detailed description of the fields included in the SWE-bench-Live dataset, along with how they are obtained during the curation process.

\begin{table}[h]
    \centering
    \caption{The required fields for a typical issue-solving task instance. Fields marked with * are \textbf{newly added} in SWE-bench-Live compared to SWE-bench.}
    \resizebox{\textwidth}{!}{
    \begin{tabular}{c|c|p{10cm}}
    \toprule
      \texttt{Field} & \texttt{Type} & Description\\
    \midrule
      \texttt{base\_commit} & \texttt{str} & The commit on which the pull request is based, representing the repository state before the issue is resolved. \\
      \texttt{patch} & \texttt{str} & Gold patch proposed by the pull request, in \texttt{.diff} format. \\
      \texttt{test\_patch} & \texttt{str} & Modifications to the test suite proposed by the pull request that are typically used to check whether the issue has been resolved. \\
      \texttt{problem\_statement} & \texttt{str} & Issue description text, typically describing the bug or requested feature, used as the task problem statement.\\
      \texttt{FAIL\_TO\_PASS} & \texttt{List[str]} & Test cases that are expected to successfully transition from failing to passing are used to evaluate the correctness of the patch.\\
      \texttt{PASS\_TO\_PASS} & \texttt{List[str]} & Test cases that are already passing prior to applying the gold patch. A correct patch shouldn't introduce regression failures in these tests.\\
      \texttt{*image\_key} & \texttt{str} & Instance-level docker image that provides an execution environment. \\
      \texttt{*test\_cmds} & \texttt{List[str]} & The command(s) used to run the test suite is identified by the verify agent in \method{}. It is required to enable detailed logging of each test item's status (e.g., by using the \texttt{pytest -rA} option).  \\
      \texttt{*log\_parser} & \texttt{str} & The type of log parser required for the instance—by default, \texttt{pytest}.\\
    \bottomrule
    \end{tabular}
    }
    \label{tab:fields}
\end{table}

\section{Experimental Setup Details}
\label{sec:experimental_setup}

In this section, we present additional details of the experimental setup to facilitate reproducibility.

\paragraph{Hyperparameters used in the experiments.} For OpenHands, we set a maximum of 60 iterations per instance, with the LLM configured to use a temperature of 0.0 and a top-p value of 1.0 as default. For SWE-agent, we limit the number of LLM calls to 100 per instance, with the temperature set to 0.0 and top-p to 1.0. For Agentless, both the number of localization samples and repair samples are set to 1, corresponding to a single rollout. The LLM temperature is set to 0.8 during the localization phase, as defined by the agent’s default, and 0.0 for all other phases. In our experiments, we omit the regression test-based reranking stage of Agentless, retaining only the localization and repair stages. The LLM calls within \method{} are configured with a temperature of 0.0.

\paragraph{Random seed in subset splitting.} The only stochastic component in this work arises during the sampling of the lite subset, where we set the random seed to 42.

\paragraph{Computational resources.} All LLM calls in this work are made through official APIs. The experiments involve parallel execution of multiple Docker containers for test execution. We conduct all the experiments on a CPU server equipped with an Intel Xeon Gold 6338 @ 2.00GHz (128 cores) and 2TB of RAM.

\section{Limitations}
\label{sec:limitations}
\textbf{Randomness caused by LLMs}: We use the LLMs as the core engine to conduct all the experiments, which might lead to potential randomness caused by different LLM calls. Since the experiments require extensive LLM calls while the overall budget is limited, we do not repeat the experiments for multiple times. To reduce the randomness, we control the execution environment to be the same and set the temperature and top\_p to zero. 

\textbf{Language limitation}: Our benchmark \benchmark primarily focuses on the Python language only, which might be limited. Since our key contribution is to propose a live benchmark with an automated and scalable method, we follow the same language choice as  existing benchmarks like SWE-bench. In the future, we plan to extend \benchmark to multiple languages such as Java, Go, and etc.

\section{Prompts in \method{}}

\label{appdix:launchprompt}

\begin{tcolorbox}[colback=gray!5,colframe=black!50,fontupper=\ttfamily,title=Prompt for Relevant Files Identification]
Given this repository structure:

------ BEGIN REPOSITORY STRUCTURE ------

\{structure\}

------ END REPOSITORY STRUCTURE ------

List the most relevant files for setting up a development environment, including:

0. CI/CD configuration files

1. README files

2. Documentation

3. Installation guides

4. Development setup guides

Format each file with its relative path (relative to project root) to be wrapped with tag <file> </file>, one per line.
\end{tcolorbox}

\begin{tcolorbox}[colback=gray!5,colframe=black!50,fontupper=\ttfamily,title=Prompt for Setup Agent]
You are a developer. Your task is to install dependencies and set up a environment that is able to run the tests of the project.\\

- You start with an initial Docker container based on \{base\_image\}.

- You interact with a Bash session inside this container.

- Project files are located in /testbed within the container, and your current working directory of bash is already set to /testbed.

- No need to clone the project again.\\

The final objective is to successfully run the tests of the project.
\\
\#\#\# Attention:

- For Python project, you should make sure the package is installed from source in the editable mode before running tests (for example 'pip install -e .') to have a development environment.

- For Python project, avoid use tox to run test if possible as it is designed specifically for CI. Read tox.ini file to find how to setup and run the test.

You run in a loop of Thought, Action, Observation.
At the end of the loop you should use Action to stop the loop.

Use Thought to describe your thoughts about the question you have been asked.

Use Action to run one of the actions available to you.

Observation will be the result of running those actions.

> Important Note: Each step, reply with only **one** (Thought, Action) pair.

> Important Note: Do not reply **Observation**, it will be provided by the system.

Your available actions are:

\{tools\}

Observation will be the result of running those actions.\\

Project Structure: the structure of the project, including files and directories.

Related Files: the content of related files of the project that may help you understand the project.

Thought: you should always think about what to do

Action: decide an action to take

Observation: the result of the action
\\

... (this Thought/Action/Observation can repeat N times) ...
\\

Thought: I think the setup should be fine

Action: stop the setup

Answer: the final result\\

Begin

Project Structure: \{project\_structure\}

Related Files: \{docs\}
\end{tcolorbox}

\begin{tcolorbox}[colback=gray!5,colframe=black!50,fontupper=\ttfamily,title=Prompt for Verify Agent]
You are a developer. Your task is to verify whether the environment for the given project is set up correctly. Your colleague has set up a Docker environment for the project. You need to verify if it can successfully run the tests of the project.\\

- You interact with a Bash session inside this container.

- The container is based on \{base\_image\}.

- The setup commands that your colleague has run are \{setup\_commands\}

- Project files are located in /testbed within the container, and your current working directory of bash is already set to /testbed.

- Use the same test framework as your colleague, because that aligns with the setup stage.

- Only test commands, skip linting/packaging/publishing commands.

- Do not change the state of the environment, your task is to verify not to fix it. If you see issues, report it not fix it.

- You can tolerate a few test cases failures—as long as most tests pass, it's good enough. \\

\#\# Important Note: \\

Your test command must output detailed pass/fail status for each test item. This is mandatory. For example, with pytest, use the -rA option to get output like: \\

```

PASSED tests/test\_resources.py::test\_fetch\_centromeres

PASSED tests/test\_vis.py::test\_to\_ucsc\_colorstring

```\\

Since we need to parse the test output to extract a test item → status mapping, **this requirement is mandatory**. If you observed that your test command does not produce such detailed output, you must adjust it accordingly. \\

In summary, your goal is:

1. Write the test commands that could output detailed pass/fail status for each test item, you can iterate until it does. (this is mandatory, DO NOT ignore this requirement!!! This is your obligation to correctly identify the test commands to run the test suite of the project, and find a way to output detailed pass/fail status)

2. Run the test command to verify if the environment is set up correctly. If not, report any observed issues. If you think the setup is correct, report none issue.
\end{tcolorbox}

\begin{tcolorbox}[colback=gray!5,colframe=black!50,fontupper=\ttfamily,title=Prompt for Base Image Selection]
Based on related file:

\{related\_files\}

Please recommend a suitable base Docker image. Consider:\\

1. The programming language and version requirements

2. Common system dependencies

3. Use official images when possible \\

Select a base image from the following candidate list:
\{candidate\_images\}

Wrap the image name in a block like <image>python:3.11</image> to indicate your choice.
\end{tcolorbox}

\section{NeurIPS Paper Checklist}


\begin{enumerate}

\item {\bf Claims}
    \item[] Question: Do the main claims made in the abstract and introduction accurately reflect the paper's contributions and scope?
    \item[] Answer:  \answerYes{} 
    \item[] Justification: For detailed contributions and scope, please see Section~\ref{sec:method} and Section~\ref{sec:experiments}
    \item[] Guidelines:
    \begin{itemize}
        \item The answer NA means that the abstract and introduction do not include the claims made in the paper.
        \item The abstract and/or introduction should clearly state the claims made, including the contributions made in the paper and important assumptions and limitations. A No or NA answer to this question will not be perceived well by the reviewers. 
        \item The claims made should match theoretical and experimental results, and reflect how much the results can be expected to generalize to other settings. 
        \item It is fine to include aspirational goals as motivation as long as it is clear that these goals are not attained by the paper. 
    \end{itemize}

\item {\bf Limitations}
    \item[] Question: Does the paper discuss the limitations of the work performed by the authors?
    \item[] Answer: \answerYes{} 
    \item[] Justification: For details about the limitation, please see Section~\ref{sec:limitations} 
    \item[] Guidelines:
    \begin{itemize}
        \item The answer NA means that the paper has no limitation while the answer No means that the paper has limitations, but those are not discussed in the paper. 
        \item The authors are encouraged to create a separate "Limitations" section in their paper.
        \item The paper should point out any strong assumptions and how robust the results are to violations of these assumptions (e.g., independence assumptions, noiseless settings, model well-specification, asymptotic approximations only holding locally). The authors should reflect on how these assumptions might be violated in practice and what the implications would be.
        \item The authors should reflect on the scope of the claims made, e.g., if the approach was only tested on a few datasets or with a few runs. In general, empirical results often depend on implicit assumptions, which should be articulated.
        \item The authors should reflect on the factors that influence the performance of the approach. For example, a facial recognition algorithm may perform poorly when image resolution is low or images are taken in low lighting. Or a speech-to-text system might not be used reliably to provide closed captions for online lectures because it fails to handle technical jargon.
        \item The authors should discuss the computational efficiency of the proposed algorithms and how they scale with dataset size.
        \item If applicable, the authors should discuss possible limitations of their approach to address problems of privacy and fairness.
        \item While the authors might fear that complete honesty about limitations might be used by reviewers as grounds for rejection, a worse outcome might be that reviewers discover limitations that aren't acknowledged in the paper. The authors should use their best judgment and recognize that individual actions in favor of transparency play an important role in developing norms that preserve the integrity of the community. Reviewers will be specifically instructed to not penalize honesty concerning limitations.
    \end{itemize}

\item {\bf Theory assumptions and proofs}
    \item[] Question: For each theoretical result, does the paper provide the full set of assumptions and a complete (and correct) proof?
    \item[] Answer: \answerNA{} 
    \item[] Justification: Not applicable as the paper is about dataset and experimental analysis rather than a theory paper. 
    \item[] Guidelines:
    \begin{itemize}
        \item The answer NA means that the paper does not include theoretical results. 
        \item All the theorems, formulas, and proofs in the paper should be numbered and cross-referenced.
        \item All assumptions should be clearly stated or referenced in the statement of any theorems.
        \item The proofs can either appear in the main paper or the supplemental material, but if they appear in the supplemental material, the authors are encouraged to provide a short proof sketch to provide intuition. 
        \item Inversely, any informal proof provided in the core of the paper should be complemented by formal proofs provided in appendix or supplemental material.
        \item Theorems and Lemmas that the proof relies upon should be properly referenced. 
    \end{itemize}

    \item {\bf Experimental result reproducibility}
    \item[] Question: Does the paper fully disclose all the information needed to reproduce the main experimental results of the paper to the extent that it affects the main claims and/or conclusions of the paper (regardless of whether the code and data are provided or not)?
    \item[] Answer: \answerYes{} 
    \item[] Justification: We present all the step-wise details in Section~\ref{sec:method} and experimental settings in Section~\ref{sec:experiments}. Besides, we  release all the code, data, and Docker environment for researchers to easily reproduce the results.
    \item[] Guidelines:
    \begin{itemize}
        \item The answer NA means that the paper does not include experiments.
        \item If the paper includes experiments, a No answer to this question will not be perceived well by the reviewers: Making the paper reproducible is important, regardless of whether the code and data are provided or not.
        \item If the contribution is a dataset and/or model, the authors should describe the steps taken to make their results reproducible or verifiable. 
        \item Depending on the contribution, reproducibility can be accomplished in various ways. For example, if the contribution is a novel architecture, describing the architecture fully might suffice, or if the contribution is a specific model and empirical evaluation, it may be necessary to either make it possible for others to replicate the model with the same dataset, or provide access to the model. In general. releasing code and data is often one good way to accomplish this, but reproducibility can also be provided via detailed instructions for how to replicate the results, access to a hosted model (e.g., in the case of a large language model), releasing of a model checkpoint, or other means that are appropriate to the research performed.
        \item While NeurIPS does not require releasing code, the conference does require all submissions to provide some reasonable avenue for reproducibility, which may depend on the nature of the contribution. For example
        \begin{enumerate}
            \item If the contribution is primarily a new algorithm, the paper should make it clear how to reproduce that algorithm.
            \item If the contribution is primarily a new model architecture, the paper should describe the architecture clearly and fully.
            \item If the contribution is a new model (e.g., a large language model), then there should either be a way to access this model for reproducing the results or a way to reproduce the model (e.g., with an open-source dataset or instructions for how to construct the dataset).
            \item We recognize that reproducibility may be tricky in some cases, in which case authors are welcome to describe the particular way they provide for reproducibility. In the case of closed-source models, it may be that access to the model is limited in some way (e.g., to registered users), but it should be possible for other researchers to have some path to reproducing or verifying the results.
        \end{enumerate}
    \end{itemize}

\item {\bf Open access to data and code}
    \item[] Question: Does the paper provide open access to the data and code, with sufficient instructions to faithfully reproduce the main experimental results, as described in supplemental material?
    \item[] Answer: \answerYes{} 
    \item[] Justification: All the code and data are open accessible with proper README instructions.
    \item[] Guidelines:
    \begin{itemize}
        \item The answer NA means that paper does not include experiments requiring code.
        \item Please see the NeurIPS code and data submission guidelines (\url{https://nips.cc/public/guides/CodeSubmissionPolicy}) for more details.
        \item While we encourage the release of code and data, we understand that this might not be possible, so “No” is an acceptable answer. Papers cannot be rejected simply for not including code, unless this is central to the contribution (e.g., for a new open-source benchmark).
        \item The instructions should contain the exact command and environment needed to run to reproduce the results. See the NeurIPS code and data submission guidelines (\url{https://nips.cc/public/guides/CodeSubmissionPolicy}) for more details.
        \item The authors should provide instructions on data access and preparation, including how to access the raw data, preprocessed data, intermediate data, and generated data, etc.
        \item The authors should provide scripts to reproduce all experimental results for the new proposed method and baselines. If only a subset of experiments are reproducible, they should state which ones are omitted from the script and why.
        \item At submission time, to preserve anonymity, the authors should release anonymized versions (if applicable).
        \item Providing as much information as possible in supplemental material (appended to the paper) is recommended, but including URLs to data and code is permitted.
    \end{itemize}

\item {\bf Experimental setting/details}
    \item[] Question: Does the paper specify all the training and test details (e.g., data splits, hyperparameters, how they were chosen, type of optimizer, etc.) necessary to understand the results?
    \item[] Answer:\answerYes{} 
    \item[] Justification: We presented the experimental details in Appendix~\ref{sec:experimental_setup} 
    \item[] Guidelines:
    \begin{itemize}
        \item The answer NA means that the paper does not include experiments.
        \item The experimental setting should be presented in the core of the paper to a level of detail that is necessary to appreciate the results and make sense of them.
        \item The full details can be provided either with the code, in appendix, or as supplemental material.
    \end{itemize}

\item {\bf Experiment statistical significance}
    \item[] Question: Does the paper report error bars suitably and correctly defined or other appropriate information about the statistical significance of the experiments?
    \item[] Answer: \answerNo{} 
    \item[] Justification: We do not report the experiment statistical significance due to the limited budget for repeat experiments. We have controlled the experimental randomness by setting temperature to 0.
    \item[] Guidelines:
    \begin{itemize}
        \item The answer NA means that the paper does not include experiments.
        \item The authors should answer "Yes" if the results are accompanied by error bars, confidence intervals, or statistical significance tests, at least for the experiments that support the main claims of the paper.
        \item The factors of variability that the error bars are capturing should be clearly stated (for example, train/test split, initialization, random drawing of some parameter, or overall run with given experimental conditions).
        \item The method for calculating the error bars should be explained (closed form formula, call to a library function, bootstrap, etc.)
        \item The assumptions made should be given (e.g., Normally distributed errors).
        \item It should be clear whether the error bar is the standard deviation or the standard error of the mean.
        \item It is OK to report 1-sigma error bars, but one should state it. The authors should preferably report a 2-sigma error bar than state that they have a 96\% CI, if the hypothesis of Normality of errors is not verified.
        \item For asymmetric distributions, the authors should be careful not to show in tables or figures symmetric error bars that would yield results that are out of range (e.g. negative error rates).
        \item If error bars are reported in tables or plots, The authors should explain in the text how they were calculated and reference the corresponding figures or tables in the text.
    \end{itemize}

\item {\bf Experiments compute resources}
    \item[] Question: For each experiment, does the paper provide sufficient information on the computer resources (type of compute workers, memory, time of execution) needed to reproduce the experiments?
    \item[] Answer: \answerYes{} 
    \item[] Justification: We provide details about the experimental settings in Section~\ref{sec:experiments} and Appendix~\ref{sec:experimental_setup}.
    \item[] Guidelines:
    \begin{itemize}
        \item The answer NA means that the paper does not include experiments.
        \item The paper should indicate the type of compute workers CPU or GPU, internal cluster, or cloud provider, including relevant memory and storage.
        \item The paper should provide the amount of compute required for each of the individual experimental runs as well as estimate the total compute. 
        \item The paper should disclose whether the full research project required more compute than the experiments reported in the paper (e.g., preliminary or failed experiments that didn't make it into the paper). 
    \end{itemize}
    
\item {\bf Code of ethics}
    \item[] Question: Does the research conducted in the paper conform, in every respect, with the NeurIPS Code of Ethics \url{https://neurips.cc/public/EthicsGuidelines}?
    \item[] Answer: \answerYes{} 
    \item[] Justification: We have followed all the necessary code of ethics to conduct the research 
    \item[] Guidelines:
    \begin{itemize}
        \item The answer NA means that the authors have not reviewed the NeurIPS Code of Ethics.
        \item If the authors answer No, they should explain the special circumstances that require a deviation from the Code of Ethics.
        \item The authors should make sure to preserve anonymity (e.g., if there is a special consideration due to laws or regulations in their jurisdiction).
    \end{itemize}

\item {\bf Broader impacts}
    \item[] Question: Does the paper discuss both potential positive societal impacts and negative societal impacts of the work performed?
    \item[] Answer: \answerNA{} 
    \item[] Justification: The paper mainly focuses on issue fixing, a specific software engineering task, which has no relation with societal issues.
    \item[] Guidelines:
    \begin{itemize}
        \item The answer NA means that there is no societal impact of the work performed.
        \item If the authors answer NA or No, they should explain why their work has no societal impact or why the paper does not address societal impact.
        \item Examples of negative societal impacts include potential malicious or unintended uses (e.g., disinformation, generating fake profiles, surveillance), fairness considerations (e.g., deployment of technologies that could make decisions that unfairly impact specific groups), privacy considerations, and security considerations.
        \item The conference expects that many papers will be foundational research and not tied to particular applications, let alone deployments. However, if there is a direct path to any negative applications, the authors should point it out. For example, it is legitimate to point out that an improvement in the quality of generative models could be used to generate deepfakes for disinformation. On the other hand, it is not needed to point out that a generic algorithm for optimizing neural networks could enable people to train models that generate Deepfakes faster.
        \item The authors should consider possible harms that could arise when the technology is being used as intended and functioning correctly, harms that could arise when the technology is being used as intended but gives incorrect results, and harms following from (intentional or unintentional) misuse of the technology.
        \item If there are negative societal impacts, the authors could also discuss possible mitigation strategies (e.g., gated release of models, providing defenses in addition to attacks, mechanisms for monitoring misuse, mechanisms to monitor how a system learns from feedback over time, improving the efficiency and accessibility of ML).
    \end{itemize}
    
\item {\bf Safeguards}
    \item[] Question: Does the paper describe safeguards that have been put in place for responsible release of data or models that have a high risk for misuse (e.g., pretrained language models, image generators, or scraped datasets)?
    \item[] Answer: \answerNA{} 
    \item[] Justification: We do not release models. The benchmark are about real-world issue resolving and we provided Docker containers as the isolated environment, thereby eliminating the safety risks.
    \item[] Guidelines:
    \begin{itemize}
        \item The answer NA means that the paper poses no such risks.
        \item Released models that have a high risk for misuse or dual-use should be released with necessary safeguards to allow for controlled use of the model, for example by requiring that users adhere to usage guidelines or restrictions to access the model or implementing safety filters. 
        \item Datasets that have been scraped from the Internet could pose safety risks. The authors should describe how they avoided releasing unsafe images.
        \item We recognize that providing effective safeguards is challenging, and many papers do not require this, but we encourage authors to take this into account and make a best faith effort.
    \end{itemize}

\item {\bf Licenses for existing assets}
    \item[] Question: Are the creators or original owners of assets (e.g., code, data, models), used in the paper, properly credited and are the license and terms of use explicitly mentioned and properly respected?
    \item[] Answer: \answerYes{} 
    \item[] Justification: All the code, data or models used in this work are properly cited as References.
    \item[] Guidelines:
    \begin{itemize}
        \item The answer NA means that the paper does not use existing assets.
        \item The authors should cite the original paper that produced the code package or dataset.
        \item The authors should state which version of the asset is used and, if possible, include a URL.
        \item The name of the license (e.g., CC-BY 4.0) should be included for each asset.
        \item For scraped data from a particular source (e.g., website), the copyright and terms of service of that source should be provided.
        \item If assets are released, the license, copyright information, and terms of use in the package should be provided. For popular datasets, \url{paperswithcode.com/datasets} has curated licenses for some datasets. Their licensing guide can help determine the license of a dataset.
        \item For existing datasets that are re-packaged, both the original license and the license of the derived asset (if it has changed) should be provided.
        \item If this information is not available online, the authors are encouraged to reach out to the asset's creators.
    \end{itemize}

\item {\bf New assets}
    \item[] Question: Are new assets introduced in the paper well documented and is the documentation provided alongside the assets?
    \item[] Answer: \answerYes{}
    \item[] Justification: The benchmark, datasets and code are well-documented.
    \item[] Guidelines:
    \begin{itemize}
        \item The answer NA means that the paper does not release new assets.
        \item Researchers should communicate the details of the dataset/code/model as part of their submissions via structured templates. This includes details about training, license, limitations, etc. 
        \item The paper should discuss whether and how consent was obtained from people whose asset is used.
        \item At submission time, remember to anonymize your assets (if applicable). You can either create an anonymized URL or include an anonymized zip file.
    \end{itemize}

\item {\bf Crowdsourcing and research with human subjects}
    \item[] Question: For crowdsourcing experiments and research with human subjects, does the paper include the full text of instructions given to participants and screenshots, if applicable, as well as details about compensation (if any)? 
    \item[] Answer: \answerNA{} 
    \item[] Justification: the paper does not involve crowdsourcing nor research with human subjects.
    \item[] Guidelines:
    \begin{itemize}
        \item The answer NA means that the paper does not involve crowdsourcing nor research with human subjects.
        \item Including this information in the supplemental material is fine, but if the main contribution of the paper involves human subjects, then as much detail as possible should be included in the main paper. 
        \item According to the NeurIPS Code of Ethics, workers involved in data collection, curation, or other labor should be paid at least the minimum wage in the country of the data collector. 
    \end{itemize}

\item {\bf Institutional review board (IRB) approvals or equivalent for research with human subjects}
    \item[] Question: Does the paper describe potential risks incurred by study participants, whether such risks were disclosed to the subjects, and whether Institutional Review Board (IRB) approvals (or an equivalent approval/review based on the requirements of your country or institution) were obtained?
    \item[] Answer: \answerNA{}
    \item[] Justification: the paper does not involve crowdsourcing nor research with human subjects
    \item[] Guidelines:
    \begin{itemize}
        \item The answer NA means that the paper does not involve crowdsourcing nor research with human subjects.
        \item Depending on the country in which research is conducted, IRB approval (or equivalent) may be required for any human subjects research. If you obtained IRB approval, you should clearly state this in the paper. 
        \item We recognize that the procedures for this may vary significantly between institutions and locations, and we expect authors to adhere to the NeurIPS Code of Ethics and the guidelines for their institution. 
        \item For initial submissions, do not include any information that would break anonymity (if applicable), such as the institution conducting the review.
    \end{itemize}

\item {\bf Declaration of LLM usage}
    \item[] Question: Does the paper describe the usage of LLMs if it is an important, original, or non-standard component of the core methods in this research? Note that if the LLM is used only for writing, editing, or formatting purposes and does not impact the core methodology, scientific rigorousness, or originality of the research, declaration is not required.
    \item[] Answer: \answerYes{} 
    \item[] Justification: LLMs are core part in the paper, we have listed all the LLM models used in our experiments in Section~\ref{sec:experiments}.
    \item[] Guidelines:
    \begin{itemize}
        \item The answer NA means that the core method development in this research does not involve LLMs as any important, original, or non-standard components.
        \item Please refer to our LLM policy (\url{https://neurips.cc/Conferences/2025/LLM}) for what should or should not be described.
    \end{itemize}

\end{enumerate}

\end{document}